\documentclass[apj,a4]{emulateapj}

\newcommand{\Ell}{E_\parallel}      

\shorttitle{Observations of the Crab pulsar between 25 and 100\,GeV with MAGIC I}
\shortauthors{Aleksic \emph{et al.}}

\begin{document}


\title{Observations of the Crab pulsar between 25 and 100\,GeV with the MAGIC I telescope}







%
\author{
J.~Aleksi\'c\altaffilmark{1},
E.~A.~Alvarez\altaffilmark{2},
L.~A.~Antonelli\altaffilmark{3},
P.~Antoranz\altaffilmark{4},
M.~Asensio\altaffilmark{2},
M.~Backes\altaffilmark{5},
J.~A.~Barrio\altaffilmark{2},
D.~Bastieri\altaffilmark{6},
J.~Becerra Gonz\'alez\altaffilmark{7,8},
W.~Bednarek\altaffilmark{9}
% ...
}
\altaffiltext{1} {IFAE, Edifici Cn., Campus UAB, E-08193 Bellaterra, Spain}
\altaffiltext{2} {Universidad Complutense, E-28040 Madrid, Spain}
\altaffiltext{3} {INAF National Institute for Astrophysics, I-00136 Rome, Italy}
\altaffiltext{4} {Universit\`a  di Siena, and INFN Pisa, I-53100 Siena, Italy}
\altaffiltext{5} {Technische Universit\"at Dortmund, D-44221 Dortmund, Germany}
\altaffiltext{6} {Universit\`a di Padova and INFN, I-35131 Padova, Italy}
\altaffiltext{7} {Inst. de Astrof\'{\i}sica de Canarias, E-38200 La Laguna, Tenerife, Spain}
\altaffiltext{8} {Depto. de Astrof\'{\i}sica, Universidad de La Laguna, E-38206 La Laguna, Spain}
\altaffiltext{9} {University of \L\'od\'z, PL-90236 Lodz, Poland}
% ...


\begin{abstract}
 
We report on the observation of $\gamma$-rays above 25\,GeV from the Crab pulsar
 (PSR B0532+21) using the MAGIC I telescope. 
Two data sets from observations during the winter period 2007/2008 and 2008/2009 are used.
In order to discuss the spectral shape from 100\,MeV to 100\,GeV, one year of 
public {\it Fermi} Large Area Telescope ({\it Fermi}-LAT) data are also analyzed to complement the MAGIC data.
The extrapolation of the exponential cutoff spectrum determined with the
{\it Fermi}-LAT data is inconsistent with MAGIC measurements,
which requires a modification of the standard pulsar emission models. 
In the energy region between 25 and 100\,GeV,
the emission in the P1 phase (from $-0.06$ to 0.04, location of the main pulse) and the P2 phase
(from 0.32 to 0.43, location of the interpulse) can be described by power laws with spectral 
indices of $-3.1 \pm 1.0_{stat} \pm 0.3_{syst}$ and $-3.5 \pm 0.5_{stat} \pm 0.3_{syst}$, respectively. Assuming an asymmetric Lorentzian
for the pulse shape, the peak positions of the main pulse and the interpulse are estimated to be at phases
$-0.009 \pm 0.007$ and  $0.393 \pm 0.009$, while the full widths at half-maximum
are $0.025 \pm 0.008$ and  $0.053 \pm 0.015$, respectively.
\end{abstract}





\section{Introduction}

The pulsar B0531+21, also commonly known as the Crab pulsar, is the compact object left over after
a historic supernova explosion that occurred in the year 1054 AD. 
Its energetic pulsar wind creates a pulsar wind nebula, the Crab Nebula, 
an emitter of strong and steady radiation. 
The pulsar and pulsar wind nebula have been observed and studied in almost the entire accessible
 electromagnetic spectrum from about 10$^{-5}$\,eV (radio emission) to nearly 100\,TeV (very high energy $\gamma$-rays, 
e.g., \citet{HegraCrab}). The nebular emission is commonly used as a standard candle for astronomy in
various energy ranges. Recently, $\gamma$-ray flares from the Crab Nebula
were discovered in the GeV range \citep{AGILECrabFlare,FermiCrabFlare} 
and a hint of increased flux in the TeV range during a GeV flare was also reported (ATel 2921).

The Crab pulsar and several other pulsars are amongst the brightest known
 sources at 1\,GeV.
However, a spectral steepening 
made
their detection above 10\,GeV elusive despite numerous efforts 
(e.g., \citet{HESSPulsarSearch, CELESTECrab, WhippleCrabPulsar}).
The energy thresholds of imaging atmospheric Cherenkov telescopes (IACTs) were, in general,
too high, while the $\gamma$-ray collection area of satellite-borne detectors was
too small to detect pulsars above 10\,GeV.



On the other hand,
a precise measurement of the energy spectrum at and above
the steepening is an important test for the standard pulsar models,
such as the polar cap (PC), outer gap (OG), and slot gap (SG) models.

In the PC model, emission takes place
within a few neutron star (NS) radii above a PC surface
\citep{Arons1979, Daugherty1982, Daugherty1996}.
There, high energy gamma-rays above $\sim 1$\,GeV should be absorbed by a strong magnetic
 field \citep{Baring2004}, 
which results in a very sharp cutoff (so-called super-exponential cutoff) in the energy spectrum.

Extending the original idea by \citet{Arons1983},
\citet{Muslimov2003, Muslimov2004a, Muslimov2004b} and \citet{Dyks2004}
investigated the possibility of high-energy emission
along the flaring field lines at high altitudes.
This type of emission, a SG emission, can be observed at all viewing angle
and for most cases emission from the two poles can be observed.
The SG model predicts an exponential cutoff
above 1\,GeV \citep{Harding2008}.
However, such geometrically thin emission models
reproduce less than 20~\%
of the observed $\gamma$-ray fluxes of the Crab and Vela pulsars \citep{Hirotani2008}.

Seeking a different possibility of high-altitude emissions,
\citet{Cheng1986a, Cheng1986b} proposed the OG model,
hypothesizing that the emission zone is located in higher altitudes,
beyond the so-called null-charge surface.
Subsequently,
\citet{Romani1995} and \citet{Romani1996}
developed the caustic model of the OG emissions.
A three-dimensional version of such a geometrical model of OG emissions
was investigated by \citet{Cheng2000}.
Recently, \citet{Romani2010} presented an atlas of
pulse properties and proposed a method to discriminate different
emission models from the geometrical point of view.

It is noteworthy that all the existing OG or SG models
predict that the highest-energy photons are emitted via 
curvature radiation and that an exponential cutoff
appears around 10\,GeV in the spectrum of the Crab pulsar
(e.g., \citep{Tang2008}).

 

The MAGIC I telescope in its standard trigger mode has the worldwide lowest threshold of all currently
operating IACTs, around 60\,GeV.
A previous study of MAGIC data above 60\,GeV revealed a 2.9 $\sigma$ 
excess from the Crab pulsar \citep{MAGICCrab}.
Following up on this hint, we investigated an alternative trigger concept, 
the sum trigger (see section \ref{SectSumT}), which lowered the 
energy threshold of MAGIC to about 25\,GeV. 
Using this new trigger, we observed the Crab pulsar between 2007 October and 2008 February and detected high-energy $\gamma$-ray emission from the Crab
 pulsar with a significance of 6.4$\sigma$ \citep{CrabScience}. 
This detection suggests the distance of the emission region from the stellar surface to be larger than
6.2 $\pm 0.2_{stat} \pm 0.3_{syst}$ times the stellar radius, which ruled out the PC model as a viable explanation of the observed emission.
This initial detection has been briefly reported in \citet{CrabScience}. In winter 2008/2009, the Crab pulsar was observed again with MAGIC 
using the sum trigger. 

In 2008 August, the new satellite-borne $\gamma$-ray detector with 1 m$^2$ collection area, 
the {\it Fermi} Large Area Telescope ({\it Fermi}-LAT), 
became operational and measured the spectra of $\gamma$-ray pulsars up to a few tens of GeV \citep{FermiPulsarCatalogue}.
All the energy spectra are consistent with a power law with an
 exponential cutoff, though statistical uncertainties above 10\,GeV are rather large\footnote{For some of the pulsars, the phase averaged spectrum deviates from the exponential cutoff
but phase-resolved analyses revealed that the spectrum of each small pulse-phase interval is
still consistent with the exponential cutoff \citep{FermiVela,FermiXGeminga}. 
}. The cutoff energies are 
typically between 1\,GeV and 4\,GeV.
These {\it Fermi}-LAT measurements also disfavor the PC model and support the OG and the SG model \citep{FermiPulsarCatalogue}.

However, the cutoff energy of the Crab pulsar determined with the 
{\it Fermi}-LAT
 under the exponential cutoff assumption is $\sim 6$\,GeV, 
an unlikely value for the signal above 25\,GeV detected by MAGIC.
In order to verify the exponential cutoff
spectrum,
a precise comparison of the energy spectra measured 
by the two instruments is needed.
The recent detection of the Crab pulsar above 100\,GeV by the VERITAS 
Collaboration has shown that indeed the energy spectrum above the break
 is not consistent with an exponential cutoff but that it is better described
by a broken power law \citep{VeritasArxiv}. It is, however, not clear
whether the spectrum continues as a power law after the break or
there is another component above 100\,GeV 
in addition to the exponential cutoff spectrum 
because of missing flux measurements in the intermediate energy range from
25\,GeV to 100\,GeV.

The main objectives of this paper are the evaluation of the exponential cutoff
spectrum of the Crab pulsar with the MAGIC data
 and the presentation of its energy spectrum between 25\,GeV and 100\,GeV. 
We also give details of the MAGIC observations, the data selection, the analysis
 and physics results. 
We report, for the first time, separate energy spectra and a pulse profile analysis above 25\,GeV 
for both the main pulse and the interpulse. 
The large majority of the results presented in this paper are extracted from the PhD thesis
of Takayuki Saito \citep{TakaThesis}.
The paper has the following structure: 
After describing the MAGIC telescope and the sum trigger in Section 2,
we present the observation and details of the data processing in Section 3.
The detection of pulsed emission is described in Section 4. Based on the MAGIC detection
and the {\it Fermi}-LAT measurements, the evaluation of the exponential cutoff assumption is
reported in Section 5. Energy spectra for the main pulse, the interpulse, and the summed pulsed emission are presented in Section 6, followed by a discussion of the pulse profile in Section 7. We conclude the paper in Section 8 together with a theoretical interpretation of the spectrum and 
an outlook on what can be expected in the near future.

\section{The MAGIC I Telescope \label{magic}}
\subsection{System Overview}

The MAGIC I telescope is a new generation IACT located on the Canary island of
 La Palma (27$\fdg$8 N, 17$\fdg$8 W, 2225 m asl). 
 Part of the Cherenkov photons emitted by charged particles in an air shower are collected 
by a parabolic reflector
 with a diameter of 17~m and focused onto a fine pixelized camera, 
providing an image of the air shower.
The camera comprises 577 photomultiplier (PMTs) and has a field of view of $\sim 3\fdg6$ diameter. 

The fast analog PMT signals are transported via 
optical fibers to the counting house where 
the signals  are processed by the trigger system
and recorded by the data acquisition (DAQ) system. 
The trigger is normally derived from the pixels in the innermost camera
 area (the trigger area) of around 1$^\circ$ radius (325 pixels).
Each signal from the pixels in the trigger area is amplified and split 
into two signals equal in amplitude.
The signals are routed to the trigger system and to the DAQ system, respectively. 
Signals from the non-trigger pixels enter the DAQ system directly.
The trigger criteria of the standard digital trigger (hereafter called standard trigger) system
are applied in two steps: each optical signal from the trigger area is converted to an electric one and
 examined
by a discriminator with a computer-controlled threshold level; the threshold level is typically
$6-7$ photoelectrons (PhEs). 
The digital signals are then processed by a topological pattern logic, which searches for 
a close-packed group of four compact-next-neighbor pixels firing within a time window of $\sim$ 6 ns. 

In the standard trigger concept, only signals above the preset threshold
 in four compact-next-neighbor pixels can 
generate a trigger,
 while signals below the threshold or pixels not situated closely together
cannot contribute to the decision. 
This deficiency is particularly pronounced in the case of shower images
 of a light content close to the threshold, i.e., in the interesting energy region below 60\,GeV. 
For this reason, a new trigger system called sum trigger has been developed to explore the energy region down to 
$\sim 25$\,GeV. Details of the new sum trigger will be presented in the Section \ref{SectSumT} below.

The signals entering the DAQ system are recorded by flash analog-to-digital converters (FADCs)
with a sampling rate of 2\,GHz. For each event, 50 FADC slices are recorded for all the pixels.
The details of the DAQ system are described in \citet{MUXFADC}.
For the pulsar study, the central pixel of the camera was modified to record the optical flux 
from the object under study, i.e., to measure the optical pulsations of the Crab pulsar. 
The details of the central pixel system can be found in \citet{CPIX}.
The telescope tracked the Crab position with a typical precision of 0$\fdg$02. 
In addition, 
we regularly recorded calibration and pedestal events with a frequency of 25\,Hz each.
Further details on the telescope can be found in \citet{Baixeras2004}.

\subsection{The Sum Trigger}
\label{SectSumT}

As mentioned above,
the standard trigger scheme is not very efficient below 60\,GeV
because even at 25\,GeV the image covers well over 
four pixels and the signals show a wide spread in amplitude. 
In order to improve the trigger
 efficiency just above threshold, a new technique was developed, the so-called sum-trigger method. 
The main feature of
the sum trigger is the summation of the analog pixel signals from a wider camera area,
 so-called patches, followed by a discrimination of this summed-up signal. 
There are 
24 partially overlapping patches in an annulus with inner and outer camera radii of $\sim 0\fdg25$ and $\sim 0\fdg8$ respectively. Each patch comprises 18 pixels.
The threshold level for the summed signal from a patch is an amplitude of 27 
PhEs.

\begin{figure}[t]
\centering
\includegraphics[width=0.4\textwidth]{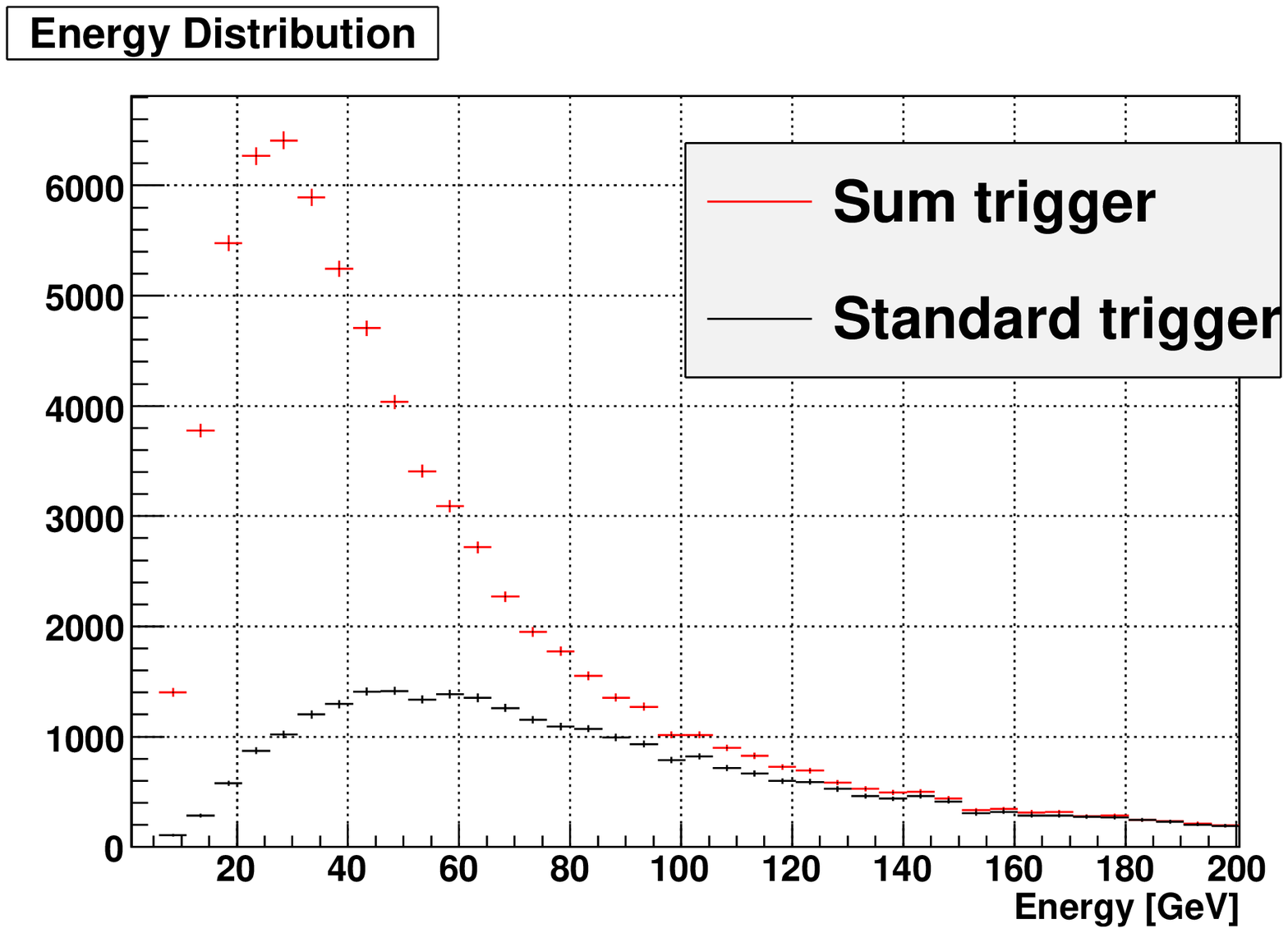}\\
\includegraphics[width=0.4\textwidth]{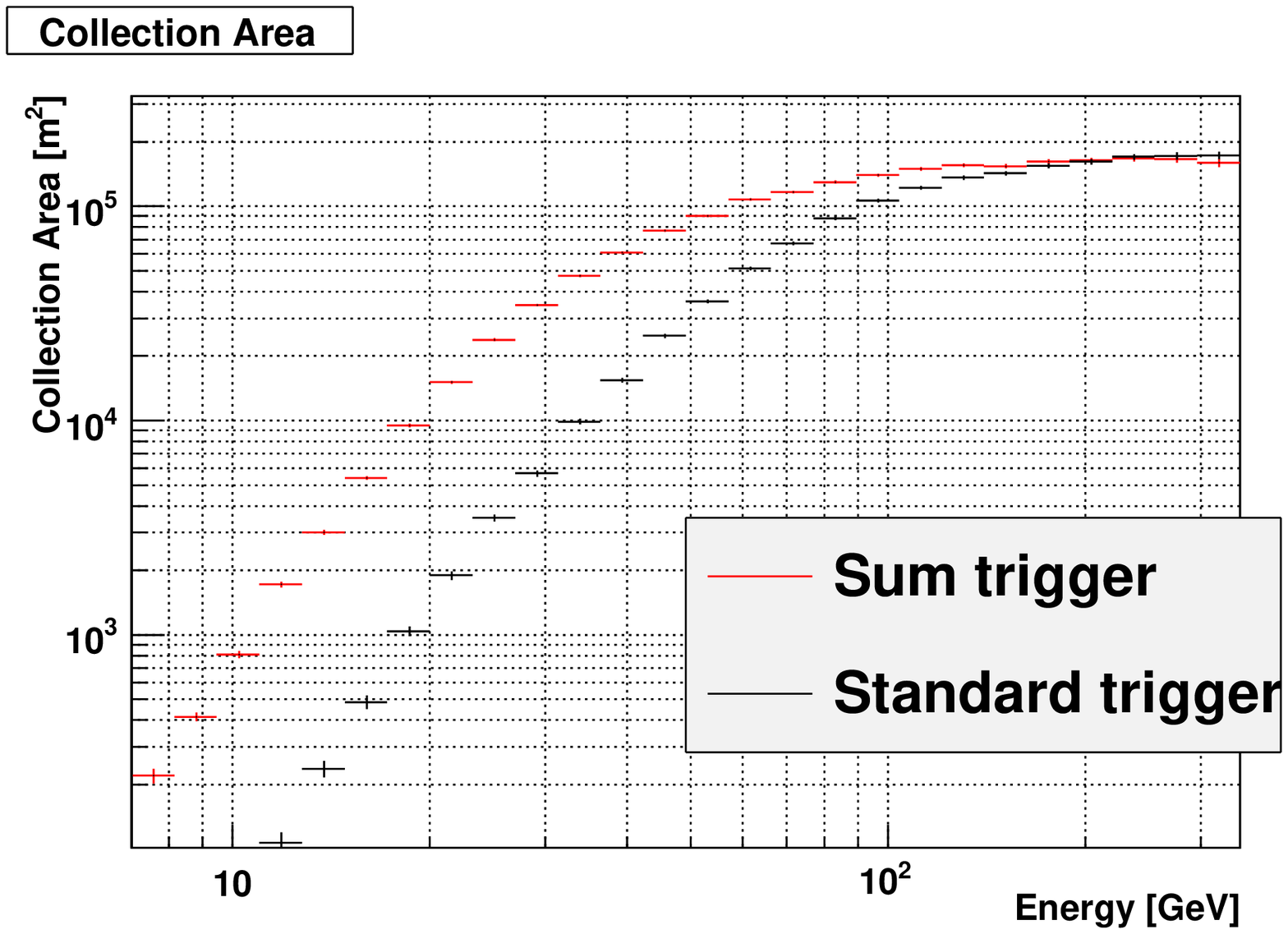}
\caption{Comparison of the energy distribution of triggered events (top)
and the gamma-ray collection area (bottom) 
for MAGIC I between the standard trigger and the sum trigger computed by MC simulations.
For the gamma-ray energy spectrum 
a power law with an index of $-2.6$ was assumed in the simulations.
}
\label{FigSumStdComp}
\end{figure}

This sum-trigger concept has some clear advantages compared to the
 standard trigger. 
The summation of the analog signals allows any pixel signal
in the patch to contribute to the trigger, 
even if its amplitude is below the pixel threshold of the standard trigger. 
The concept, however, has the disadvantages of being quite sensitive to accidental 
triggers from afterpulses.
Afterpulses are caused by PhEs hitting the dynodes and sometimes releasing ions 
from adsorbed water or adsorbed gases. These ions are back-accelerated by the relatively high voltage
between the photocathode and the first dynode, hit the photocathode and liberate many secondary electrons.
The afterpulse amplitude spectrum follows basically an exponential distribution, which drops significantly slower than the Poissonian night sky light distribution and completely dominates 
the rate of signal above 5-6 PhE. 
In a single patch, i.e., from the sum of 18 pixels, the rate of afterpulses above 27 PhEs was found to be around $20-30$\,kHz, which is far beyond the MAGIC DAQ rate limit of 1\,kHz. 
To suppress this undesirable background, individual pixel signals were, before summing, 
clipped at an amplitude of $\sim 6$ PhE, thus making the trigger insensitive to large 
afterpulses of individual PMTs. 
The sum-trigger area (the annulus with inner and outer radii of $\sim 0\fdg25$ and
 $\sim 0\fdg8$), the patch size (18 pixels), the threshold level (27 PhE in amplitude after sum)
and the clipping level ($\sim 6$ PhE) were optimized by measurements and detailed Monte Carlo (MC)
studies \citep{MichaelThesis}. 
The chosen settings for the sum trigger result in a trigger threshold of 25\,GeV for a gamma-ray source with an index of $-2.6$
as shown in the top panel of Figure \ref{FigSumStdComp}
 (according to the convention in ground-based $\gamma$-ray astronomy the threshold is defined as the peak of the reconstructed differential energy spectrum).
The sum trigger also improved significantly the collection area for low energy showers when compared to the area of the standard trigger. 
At 20\,GeV the collection area of the sum trigger is 10 times larger and at 60\,GeV still twice
as large when compared to that of the standard trigger,
as shown in the bottom panel of Figure \ref{FigSumStdComp}. 
A more detailed description of the sum trigger is presented in 
\citet{MichaelThesis} and \citet{SumT}.

\section{Observation and Data Processing \label{obs}}

\subsection{Observation}

The first observation of the Crab pulsar with the sum trigger started on 2007 October 21 and
 extended up to 2009 February 3. 
In total the Crab pulsar was observed for 48 hr in winter 2007/2008 \citep{CrabScience}
and for 78 hr in winter 2008/2009.
In the first year, all observations were restricted to zenith angles below 20$^\circ$ where
the air mass between the showers and the telescope is the lowest possible, i.e., 
 the atmospheric
 transmission for Cherenkov light is highest. In this zenith angle range 
the correlation between energy and observed
 number of PhEs is almost independent of the zenith angle and the trigger threshold
is nearly constant as a function of the zenith angle.
In the second year, some of the observations were done at zenith angles above 20$^\circ$.
These data are not used in the following analysis.
In the second campaign, five sub-patches malfunctioned. The losses are
estimated and corrected using MC simulations. 

\subsection{Data Processing}
In the calibration process, the conversion factors from the FADC counts
to the number of PhEs and the relative timing offsets of all pixels are computed
using the calibration and pedestal events. 
The details of the procedure can be found in \citet{Calibration}.
After the calibration, an image cleaning is performed 
in order to remove pixels which do not contain a useful Cherenkov photon
signal (e.g., pixels only containing FADC pedestal, NSB photons, and afterpulses).
The standard procedure of the image cleaning can be found
in \citet{Diego}. 
Since this study aims for the lowest possible threshold,
a more sophisticated method of image cleaning was used in this analysis.
At first the algorithm searches for the core pixels of the shower image.
The definitions of the core pixels are as follows.
\begin{itemize}
\item If two neighboring pixels have more than 4.7 PhE each
and the time difference is less than 0.8 ns, these two pixels are core pixels.
\item If three neighboring pixels have more than 2.7 PhE each 
and the arrival times of all three are within 0.8 ns, these three pixels are core pixels.
\item If four neighboring pixels have more than 2.0 PhE and 
the arrival times of all four are within 1.5 ns,
these four pixels are core pixels.
\end{itemize}
After the core search, boundary pixels of the image are selected.
For a pixel to be defined as a boundary pixel, the following three conditions
must be fulfilled:
\begin{enumerate}
\item The pixel must be a neighbor to at least one of the core pixels.
\item The pixel must have more than 1.4 PhE.
\item The time difference to at least one of the neighboring core pixels 
must be less than 1.0 ns.
\end{enumerate}

The charge and timing information of pixels which are 
neither ``core'' nor ``boundary'' is discarded.
In this way, accidental trigger events (half of the recorded data)
are efficiently rejected and the contamination of e.g., NSB photons
to the shower image can be mostly suppressed. Further details of this method 
can be found in \citet{Maxim}.

After the image cleaning, the conventional image parameters
are calculated in the standard way 
\citep{Hillas1985, Diego}.

\subsection{Data Pre-selection}
Only data taken under stable atmospheric and hardware conditions were used in
the analysis. Selection criteria include, for instance, the performance of the mirror
focusing and reflection, the cloud coverage and the stability of the
event rate after image cleaning. A detailed description of the
pre-selection criteria can be found in \citet{TakaThesis}. As mentioned
before, only data with zenith angles below 20$^\circ$ were used in the
analysis. After applying all criteria 25\,hr remained 
from the winter 2007/2008 period and 34\,hr from the winter 2008/2009 period.

\subsection{Event Selection}

\begin{figure}[t]
\centering
\includegraphics[width=0.4\textwidth]{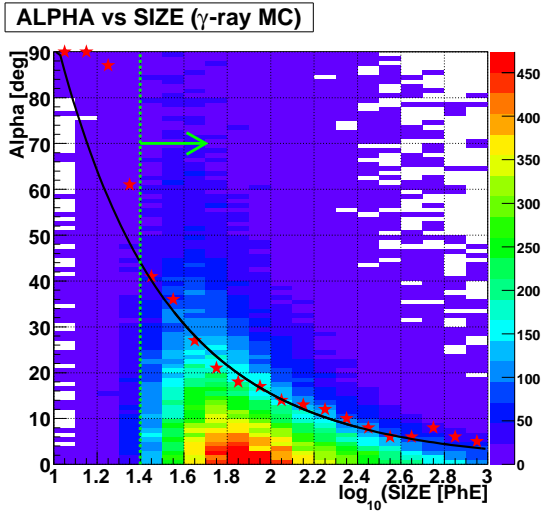}
\caption{$ALPHA$ distribution of $\gamma$-ray MC events as a function of $SIZE$.
Red stars indicate the cut values which maximize the Q-factor defined by Equation (\ref{EqQfactor}).
The black solid line shows the $SIZE$-dependent $ALPHA$ cut used in the analysis.
The green dotted line denotes $SIZE = 25$ PhE, below which data are not used for the analysis.
 }
\label{FigALPHA}
\end{figure}

A cleaned image of a 25\,GeV $\gamma$-ray on average has $\sim 8$ pixels, 
which is barely sufficient to perform a moment analysis to obtain 
the conventional Hillas image parameterization of the shower.
However, the only effective way to separate $\gamma$-rays from the background below 100\,GeV
exploits the orientation of the images in the camera plane. 
$\gamma$-rays from the source have images that point with their major axis
 towards the source location.
Background events (mainly hadron showers and muon arcs/rings), on the other hand, have images that are randomly oriented.
 The parameter that describes the orientation of an image in the camera is called $ALPHA$, 
which is the angle between the major axis of an image and the line connecting the center of gravity
(COG) of that image with the location of the source in the camera. 
The $ALPHA$ distribution of the $\gamma$-ray MC events as a function of $SIZE$ is illustrated
in Figure \ref{FigALPHA}, where $SIZE$ is the total number of PhE in the image.
Red stars indicate the cut values which maximize the so-called quality factor $Q$
defined as
\begin{eqnarray}
Q = \epsilon_\gamma / \sqrt{\epsilon_{BG}}
\label{EqQfactor}
\end{eqnarray}
 where $\epsilon_\gamma$ and $\epsilon_{BG}$ are the survival efficiencies 
of the $\gamma$-ray events and hadron background events, respectively.

\begin{figure}[t]
\centering
\includegraphics[width=0.4\textwidth]{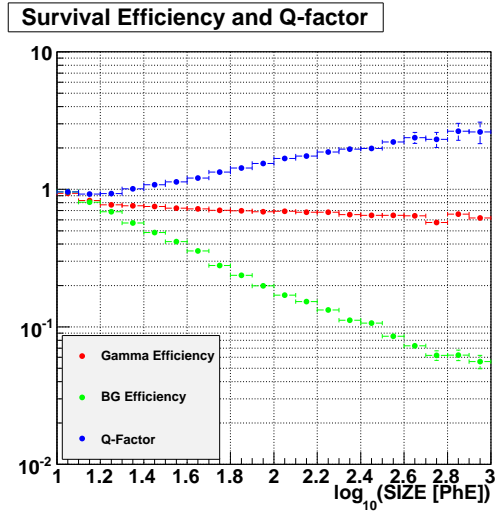}
\caption{
$\gamma$-ray survival efficiency (red), hadron background survival efficiency (green)
and the Q-factor (blue, Equation (\ref{EqQfactor})) as a function of $SIZE$ 
for the used $SIZE$-dependent $ALPHA$ cut. 
}
\label{FigQfactor}
\end{figure}

In the analysis we used an $ALPHA$ cut depending on $SIZE$, which was derived by fitting a 
numerical function $a(\log_{10}(SIZE)+b)^c$ to the best cut values found in the individual $SIZE$
 bins (stars in Figure \ref{FigALPHA}).
$\epsilon_\gamma$, $\epsilon_{BG}$ and $Q$ for the used $SIZE$-dependent $ALPHA$ cut
are shown in Figure \ref{FigQfactor}.

In addition, the so-called spark-like events, which are created by the discharge of 
charge accumulated at the glass envelope of some PMTs, are rejected.
These events are efficiently identified by applying the condition
\begin{eqnarray}
4.0\times\log_{10}(CONC)> 1.5- \log_{10}(SIZE)
\end{eqnarray}
where, $CONC$ is defined as the sum of the charges in the two pixels with the highest
content divided by $SIZE$, which indicates how strongly the charge is concentrated in 
a small region.

\subsection{Energy Estimates}
The energy of each event is estimated by means of the Random Forest method.
After the training with MC $\gamma$-ray events, the Random Forest assigns 
the most probable
 energy to each event 
by using several image parameters
in a comprehensive manner. The details of the method can be found in \citet{RF}.
In this analysis $SIZE$, $LENGTH$, $DIST$ and the zenith angle were used
for the Random Forest energy estimation. 
$LENGTH$ is the second moment of the charge distribution along 
the major axis of the shower image, 
while $DIST$ is the distance between
the source position in the camera and the center of gravity of the image. 
$WIDTH$ (the second moment along the minor axis of the image)
 and $CONC$, which are normally
included, were not used because 
it was found that they do not contribute to the energy estimate in the very low energy regime.
\\
\subsection{Pulse-phase Calculation}
Each event is marked with a time stamp, which gives the time when the event was triggered.
 The time stamps are derived from a GPS controlled Rubidium clock and have an absolute 
accuracy of less than 1\,$\mu$s.
In order to compensate the varying propagation times of the $\gamma$-rays within the solar system,
 which are mainly due to Earth's movement around the Sun, 
the recorded times are transformed to the barycenter of the solar system. 
The barycentric correction was done with the software package TEMPO [24].
The rotation frequency $\nu_0$, its time derivative $\dot \nu_0$ and
the barycentric times of the main pulse peak of the Crab pulsar 
are monitored in radio at 610\,MHz
by the Jodrell Bank radio telescope \citep{JodrellBankPaper}
and the values of the parameters are published every month\footnote{http://www.jb.man.ac.uk/~pulsar/crab.html}.
Based on these, the rotational phase of each event is computed from the barycentric times with the following formula
\begin{eqnarray}
{\rm Phase} = \nu_0 (t-t_0) + \frac{1}{2} \dot \nu_0 (t-t_0)^2
\end{eqnarray}
The second and higher derivative terms of this Taylor series are negligible on 
the scale of one month.

\begin{figure}[t]
\centering
\includegraphics[width=0.4\textwidth]{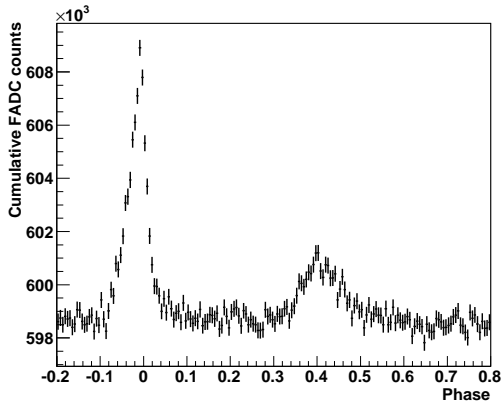}
\caption{Pulse profile of the Crab pulsar seen in optical wavelengths with 15 hr of observations. 
The signal was recorded with the central pixel of the MAGIC camera in parallel to the $\gamma$-ray measurements. }
\label{FigOptPul}
\end{figure}
\section{Pulsation of the Crab pulsar above 25\,GeV}
\subsection{Optical Pulsation}
In order to assure that the timing system of MAGIC and 
 the pulse phase calculation worked properly,
the optical pulsation of the Crab pulsar was checked first. 
The optical pulsation was measured with the central pixel, which was modified
 for this purpose to be sensitive and to digitize the light flux variation from the source.
Every time a shower event was triggered, the signal of the central pixel was recorded by the DAQ
for 25\,ns.

The phase distribution (hereafter pulse profile) of the central pixel data is shown in 
Figure \ref{FigOptPul}. Two peaks are clearly visible
at the expected phases. Phase 0 corresponds to the main peak position in radio at 610\,MHz.
 A delay of $\sim 0.01$ in phase can be seen with respect to the radio main peak
position, which
is known and consistent with other measurements (see, e.g., \citet{OosterBroek}).

\subsection{Pulsation above 25\,GeV}
\begin{figure}[t]
\centering
\includegraphics[width=0.5\textwidth]{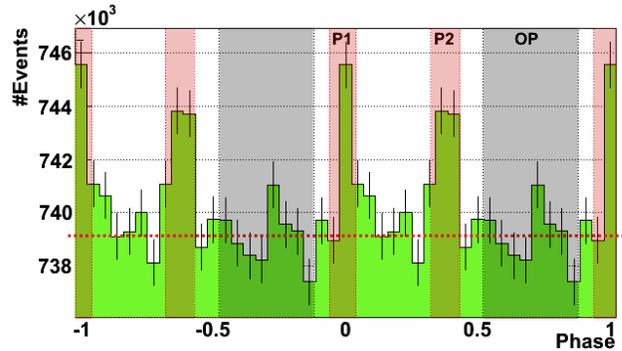}
\caption{Pulse profile of the Crab pulsar obtained with MAGIC.
The red shaded area indicates the signal phases (P1 and P2) while
the black shaded area indicates the background control phases (OP phases).
}
\label{FigLC}
\end{figure}

The pulse profile of the $\gamma$-ray events detected with MAGIC
is shown in Figure \ref{FigLC}. Events with $SIZE$ below 25\,PhE and above 500 PhE were discarded.
Note that every event is shown twice (three times for the first bin) 
in order to generate a pulse profile that spans the phase region from $-45/44$ ($-1.0227$) to $45/44$ (1.0227). 
The bin width is 1/22 in phase, which corresponds to about 1.5\,ms.
An excess is evident in the profile at the position of the main pulse and interpulse 
of the pulsar. 
Following the often-used convention \citep{Fierro1998} of 
P1 (main pulse phase from $-0.06$ to 0.04) and P2 (interpulse phases from 0.32 to 0.43), the numbers of excess events in P1 and P2
are $6200 \pm 1400$ (4.3 $\sigma$) and $11300 \pm 1500$ ($7.4\sigma$), respectively.
By summing up P1 and P2, the excess corresponds to 7.5$\sigma$. 
The background level was estimated using the so-called off pulse (OP) phases (0.52 - 0.88, \citet{Fierro1998}).
Above 25\,GeV, the flux of P2 is nearly twice that of P1. 
The width of the main pulse is significantly smaller than the conventional P1 phase interval.
By defining the signal phase of the main pulse to be $-1/44$ to 1/44 in phase (the bin with the largest number of events),
the excess is $6400 \pm 970$ corresponding to 7.0$\sigma$.
No significant emission between the main pulse and the interpulse was detected.
A detailed study of the pulse profile is given in Section \ref{SectProfile}.

\begin{figure}[t]
\centering
\includegraphics[width=0.5\textwidth]{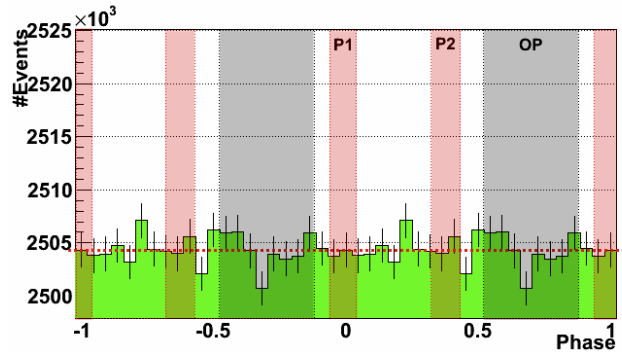}
\caption{Pulse profile of the Crab pulsar 
when the $ALPHA$ cut is inverted. As expected, the signals seen in Figure \ref{FigLC}
disappear.}
\label{FigInv}
\end{figure}

In order to verify the soundness of the signal, further tests were made.
Firstly, the phase distribution of the events which are rejected by the $ALPHA$ cut
is examined. The results are shown in Figure \ref{FigInv}. As expected, 
the distribution is consistent with statistical fluctuations of the background
without any signal. Also, the growth of the number of excess events as a function of
the number of background events is checked. As one can see in Figure \ref{FigGrowth},
the excess grows linearly, assuring that the signal is constantly detected.

\subsection{Variability Study}
\begin{figure}[t]
\centering
\includegraphics[width=0.5\textwidth]{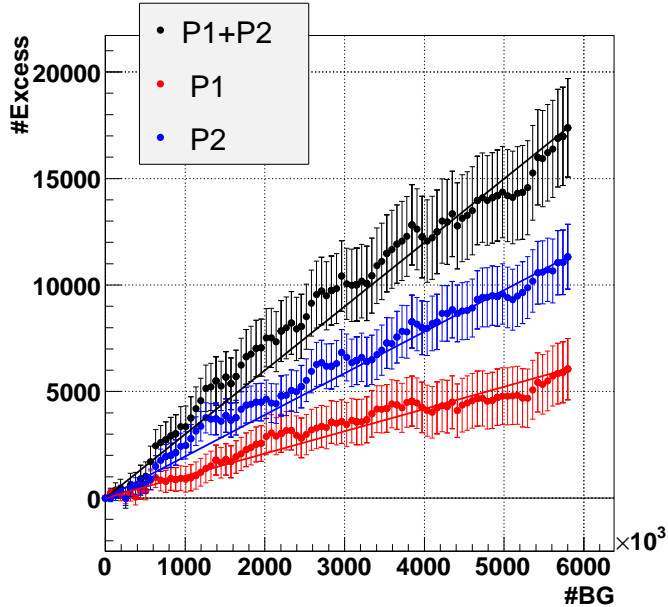}
\caption{Numbers of excess events as a function of the number of background events
(events in OP phases)
for P1 (red), P2 (blue), and the sum of the two (black).
They grow linearly, implying constant detection of the signal.
}
\label{FigGrowth}
\end{figure}

\begin{figure}[h]
\centering
\includegraphics[width=0.23\textwidth]{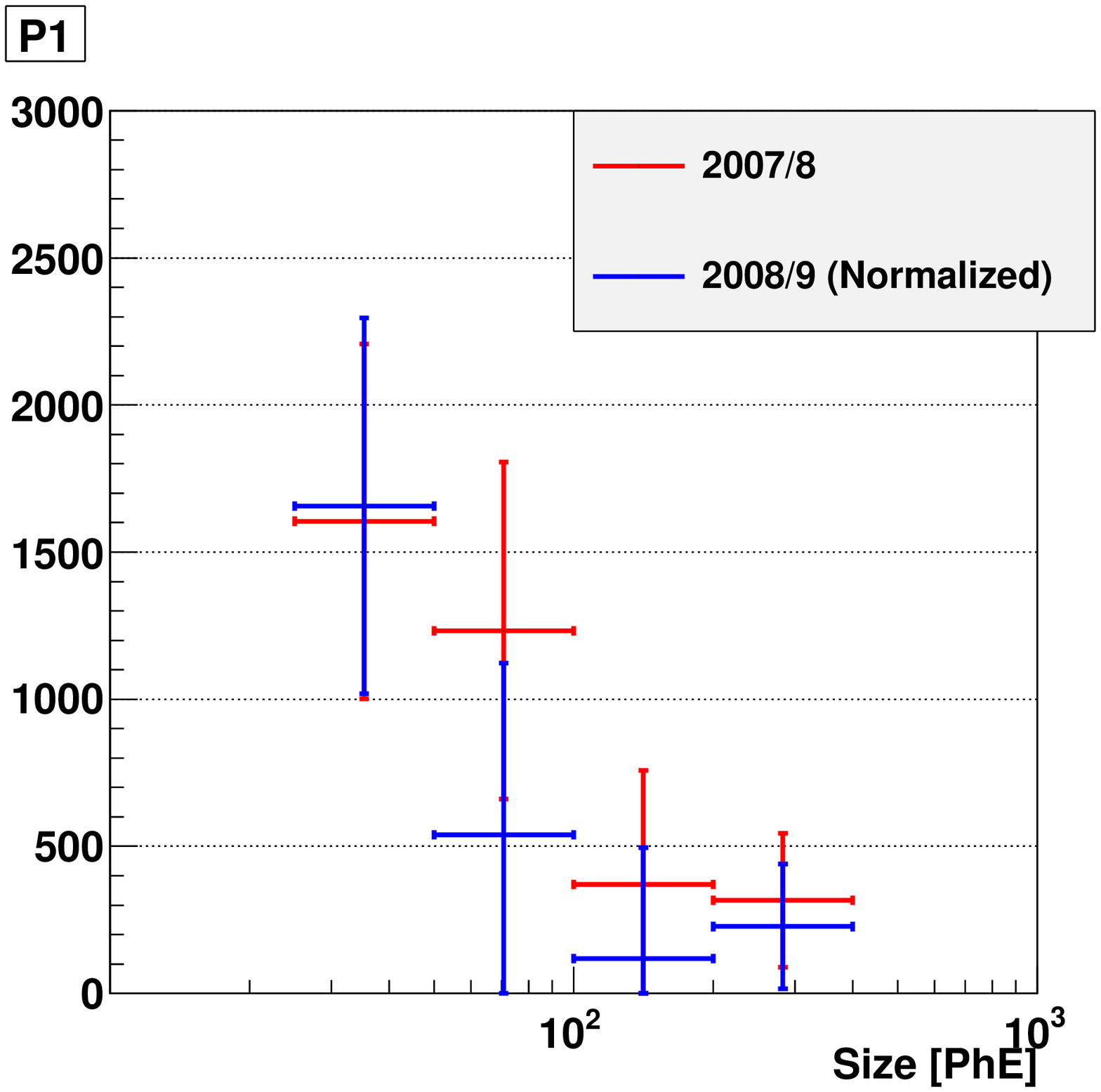}
\includegraphics[width=0.23\textwidth]{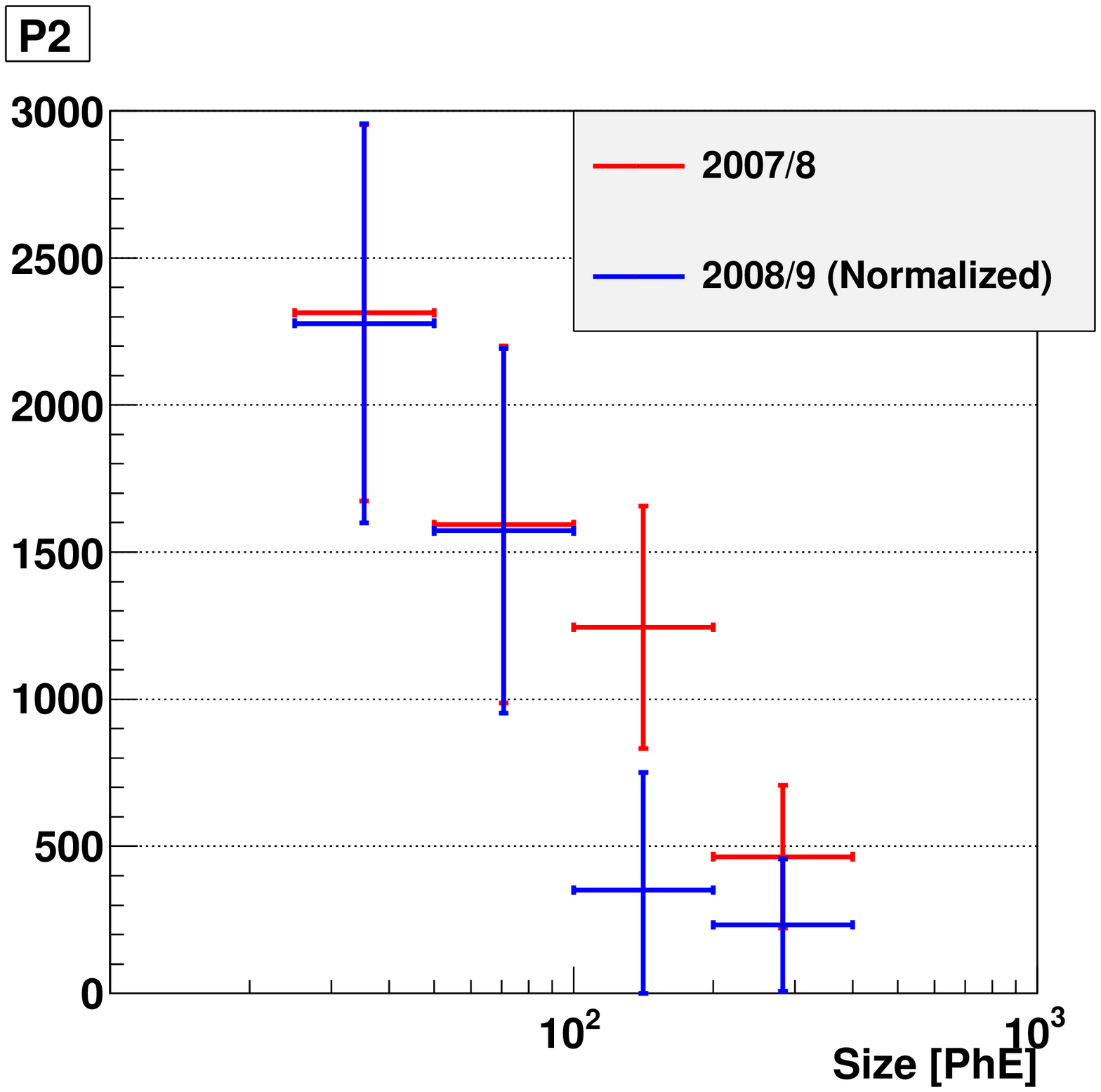}
\caption{The comparison of the $SIZE$ distributions between the winter 2007/2008 and the winter
2008/2009. The difference in observation time and the hardware performance 
are corrected by scaling the data of winter 2008/2009.
}
\label{FigExComp}
\end{figure}

The linear growth of the excess shown in Figure \ref{FigGrowth}
implies a constant flux of the pulsed signal. Nevertheless, we also applied the $\chi^2$ method
to test for a possible yearly variability.
The number of excess events as a function of $SIZE$ 
is compared between the two years in Figure \ref{FigExComp}.
The difference of observation time and the effect of the malfunctioning of
the sum-trigger sub-patches are corrected for the second year data. 
Using MC simulations, the sub-patch malfunction effects on the acceptance 
were estimated to be
 21\%, 17\%, 11\% and 7\% for the $SIZE$ ranges of 25-50, 50-100, 100-200 and
200 - 400 PhE, respectively. 
The $\chi^2$ values for the comparison of the two years are 1.0 and 3.1 for
P1 and P2, respectively, for 4 degrees of freedom.
No significant yearly variability was found in the flux.

\begin{figure}[h]
\centering
\includegraphics[width=0.4\textwidth]{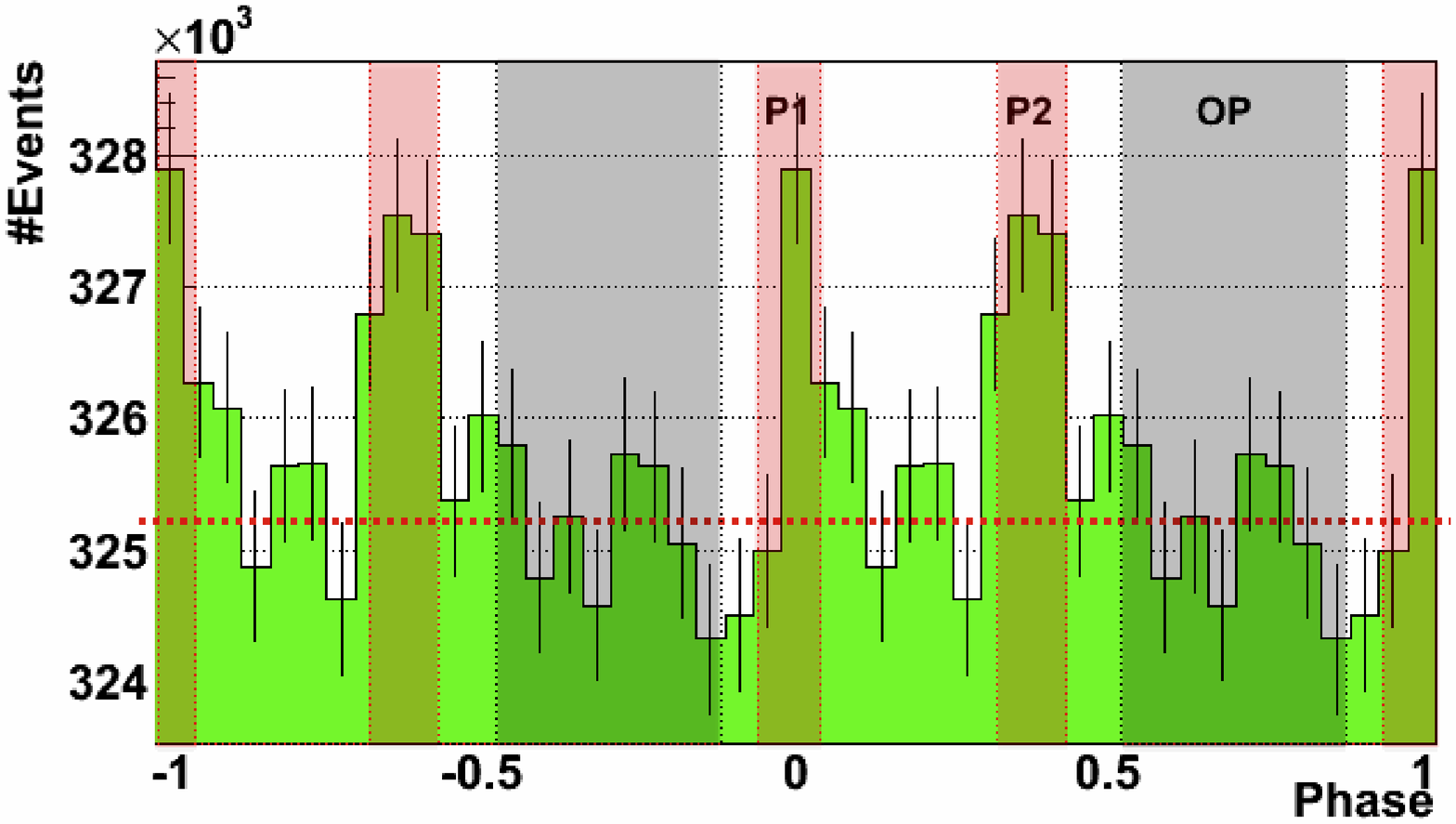}\\
\includegraphics[width=0.4\textwidth]{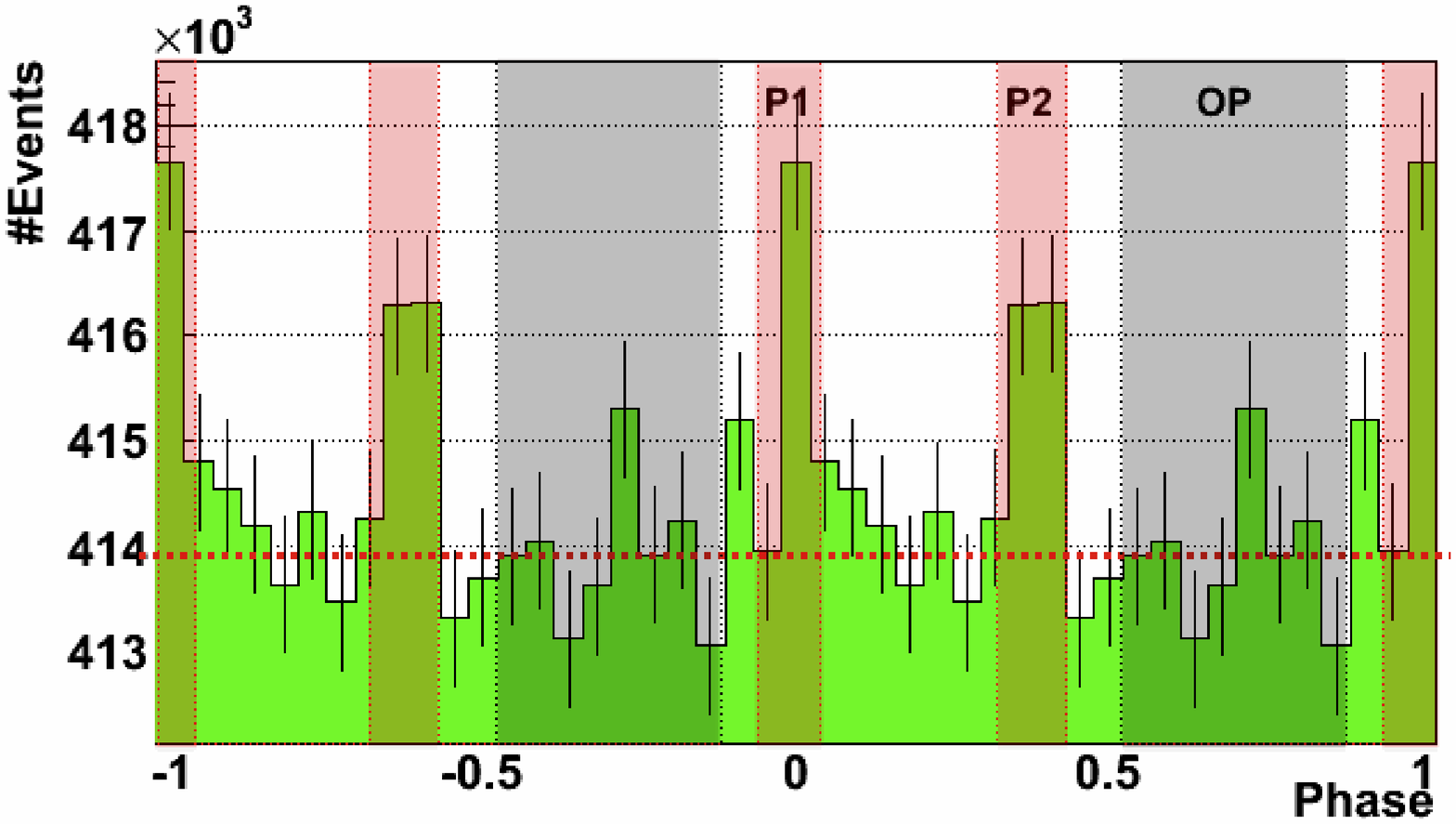}
\caption{Pulse profiles of the Crab pulsar for the winter 2007/2008 (top)
and the winter 2008/2009 (bottom). The two profiles are statistically consistent.
}
\label{FigYearlyLC}
\end{figure}

We also studied a possible variability of the pulse profile.
Figure \ref{FigYearlyLC} shows the pulse profiles for the first (winter 2007/20088)
and the second (winter 2008/2009) year. 
The two profiles are compared with each other with a $\chi^2$ test
from phase $-0.0682$ to 0.432 ($-3/44$ to 19/44), which is roughly from the beginning 
of P1 to the end of P2. The obtained $\chi^2$ is 5.0 for 10 degrees of freedom. 
Therefore, no significant yearly variability of the pulse profile was found.

\section{Evaluation of the exponential cutoff spectrum suggested from the {\it Fermi}-LAT data}

It is very important for the verification
of the standard OG and SG models to check if the energy spectrum follows
an exponential cutoff.
All energy spectra measured by {\it Fermi}-LAT up to a few tens of GeV are indeed 
consistent with the OG/SG model. 
In this section, we evaluate the exponential cutoff hypothesis based on 
 the measurements performed by {\it Fermi}-LAT and MAGIC.

\subsection{Analysis of Public {\it Fermi}-LAT Data}
\begin{figure}[t]
\centering
\includegraphics[width=0.4\textwidth]{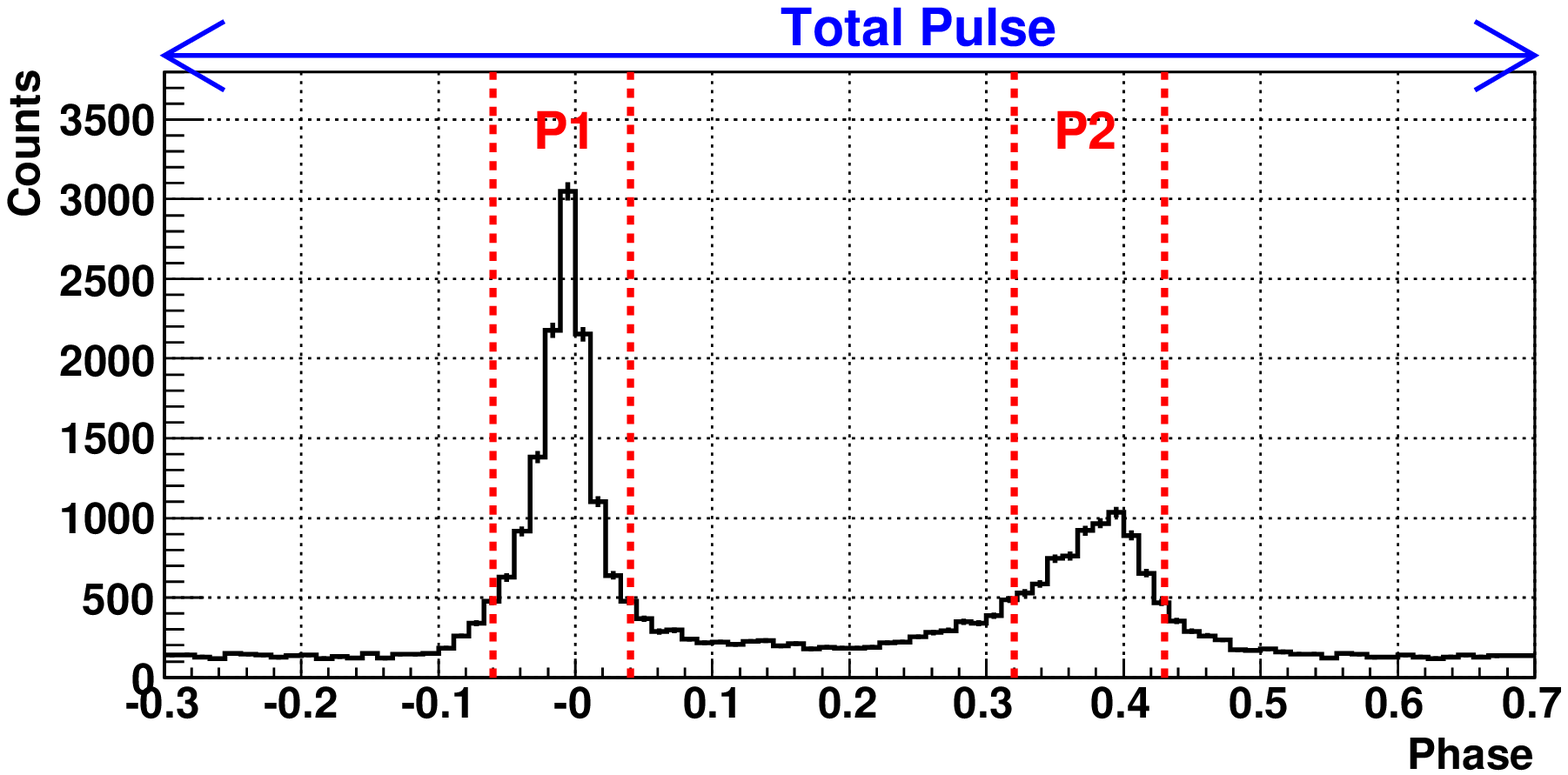}\\
\includegraphics[width=0.4\textwidth]{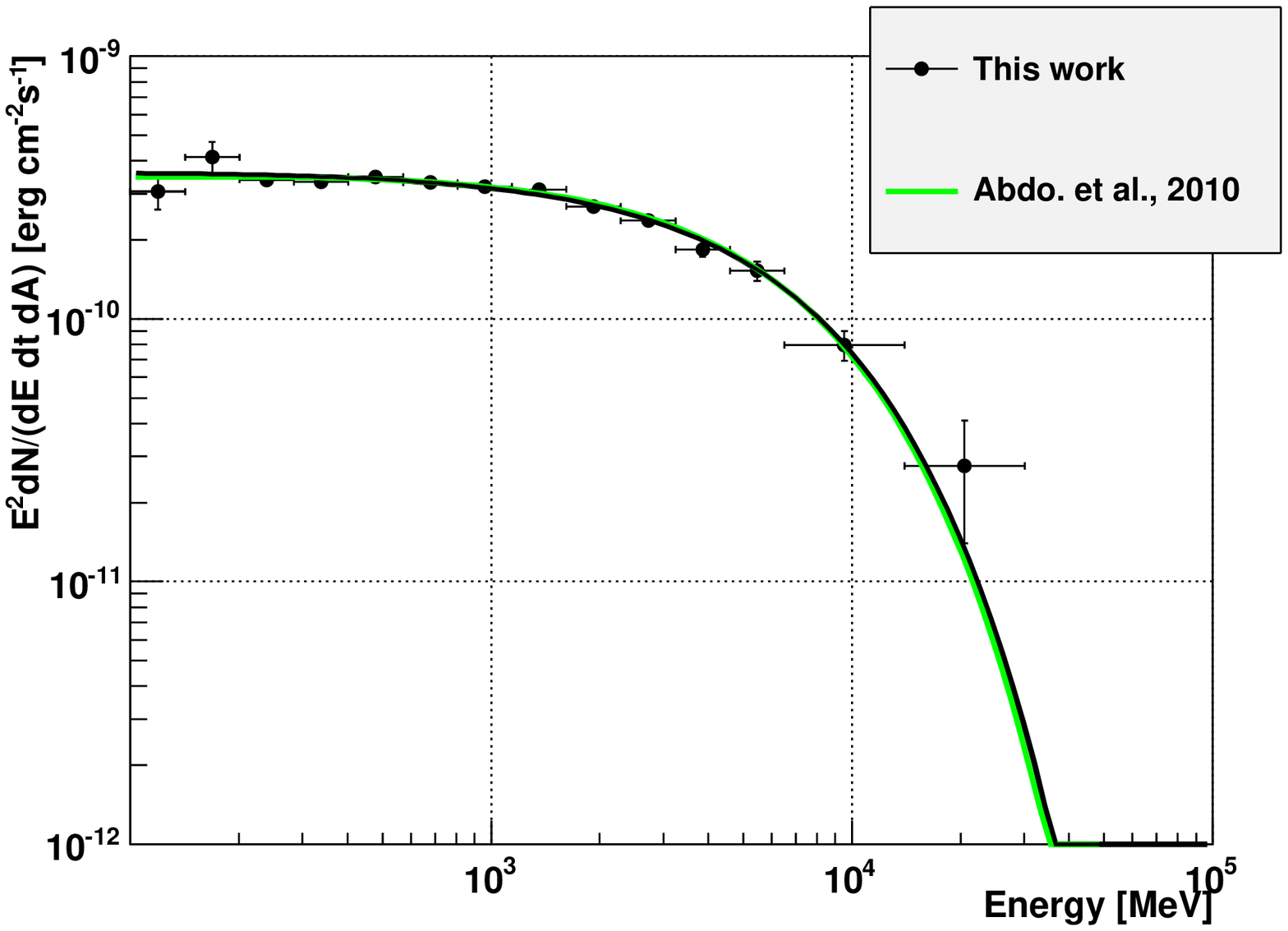}
\caption{
Top: the pulse profile of the Crab pulsar above 100\,MeV produced with the {\it Fermi}-LAT data.
Bottom: the energy spectrum of the Crab pulsar (total pulse).
The black line and dots are obtained from the public {\it Fermi}-LAT data,
while the green line is the spectrum reported in \citet{FermiCrab}.
}
\label{FigSpectrum}
\end{figure}

Although the {\it Fermi}-LAT Collaboration published their results of the Crab pulsar observations
\citep{FermiCrab}, 
we performed a customized analysis of the specific phase intervals in order to properly compare
the {\it Fermi}-LAT and MAGIC data.
One year of {\it Fermi}-LAT data taken from 2008 August 4 to 2009 August 3 are analyzed. 
Events with an energy between 100\,MeV and 300\,GeV and with an arrival direction within a radius of 20 degrees
 around the Crab pulsar were downloaded from the Fermi Science Support Center.\footnote{http://fermi.gsfc.nasa.gov/ssc/data/}
Only events with the highest probability of being photons, those in the diffuse class, were used in this
analysis. Events with imperfect spacecraft information and events taken when the satellite 
was in the South Atlantic Anomaly were rejected. In addition, a cut on the maximum zenith angle
($< 105 \degr$) was applied to reduce the contamination from the Earth-albedo $\gamma$-rays, which are produced
by cosmic rays interacting with the upper atmosphere.
The pulse-phase assignment to each event was carried out by the {\it Fermi}-LAT analysis
tool with the monthly ephemeris information from Jodrell Bank.\footnote{http://www.jb.man.ac.uk/~pulsar/crab.html}
For the detector response function, ``P6\_V3\_Diffuse'' is used, while 
``isotropic\_iem\_v02.txt'' and ``gll\_iem\_v02.fit'' are used for the extragalactic and
galactic diffuse emission models. 
 In order to estimate the contribution of the Crab Nebula,
the pulse-phase interval between 0.52 and 0.87 is used.\footnote{This phase range is not identical to
the OP phases ($0.52 - 0.88$; \citet{Fierro1998}), which was used for the MAGIC data to estimate the background level. 
We adopt this range to be consistent with \citet{FermiCrab}
}.
The contamination from nearby bright sources such as Geminga and IC 443 are also 
taken into account in the calculation of the spectrum.
 The unbinned likelihood spectral analysis assuming a power-law spectrum with an exponential cutoff 
\begin{eqnarray}
\frac{{\rm d}^3 F(E)}{{\rm d} E{\rm d}A{\rm d}t} = F_{1} (E/1{\rm ~GeV})^{-\Gamma_1}\exp(-E/E_c)
\label{EqExpCut}
\end{eqnarray}
gives $F_{1} = (2.32 \pm 0.05_{stat}) \times 10^{-10}~{\rm cm^{-2}s^{-1} MeV^{-1}}$,
$\Gamma_1~=~{1.99~\pm~0.02_{stat}}$ and $E_c = 6.1 \pm 0.5_{stat}$\,GeV as best fit parameters,
consistent with the values in \citet{FermiCrab}.
Results for the total pulse are shown by the solid black line in Figure \ref{FigSpectrum},
together with the pulse profile above 100\,MeV. 
The green curve in the figure is the spectrum given in \citet{FermiCrab}. 
 The points are obtained by applying the same likelihood method in the limited energy intervals,
assuming a power-law spectrum within each interval.

The same analysis was applied to P1, P2 and the sum of the two (P1~+~P2).
The results are shown in Figure \ref{FigSpectrumPhase} and the best-fit parameters are
summarized in Table \ref{TabSpec}.

\begin{figure}[h]
\centering
\includegraphics[width=0.395\textwidth]{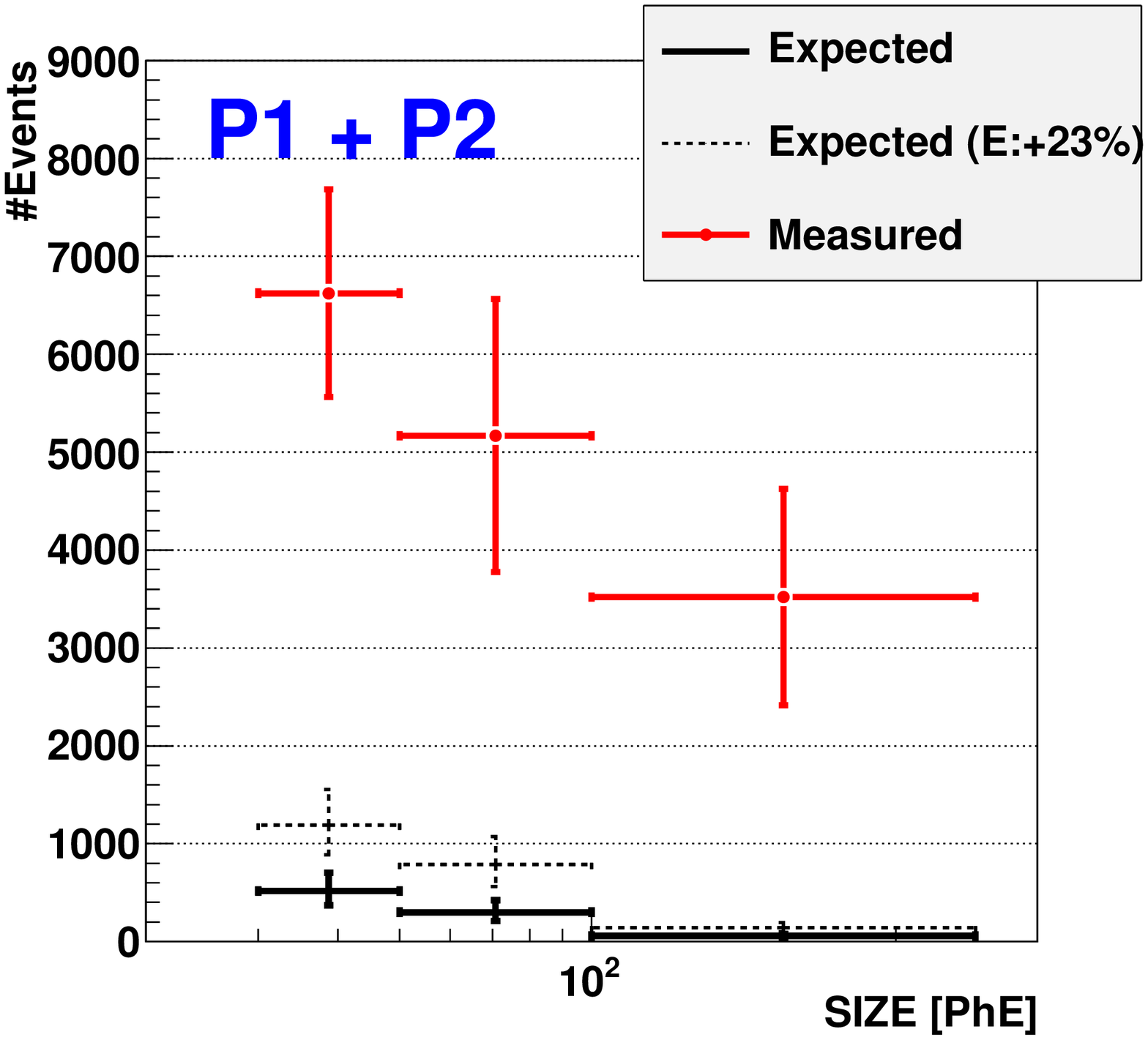}\\
\includegraphics[width=0.395\textwidth]{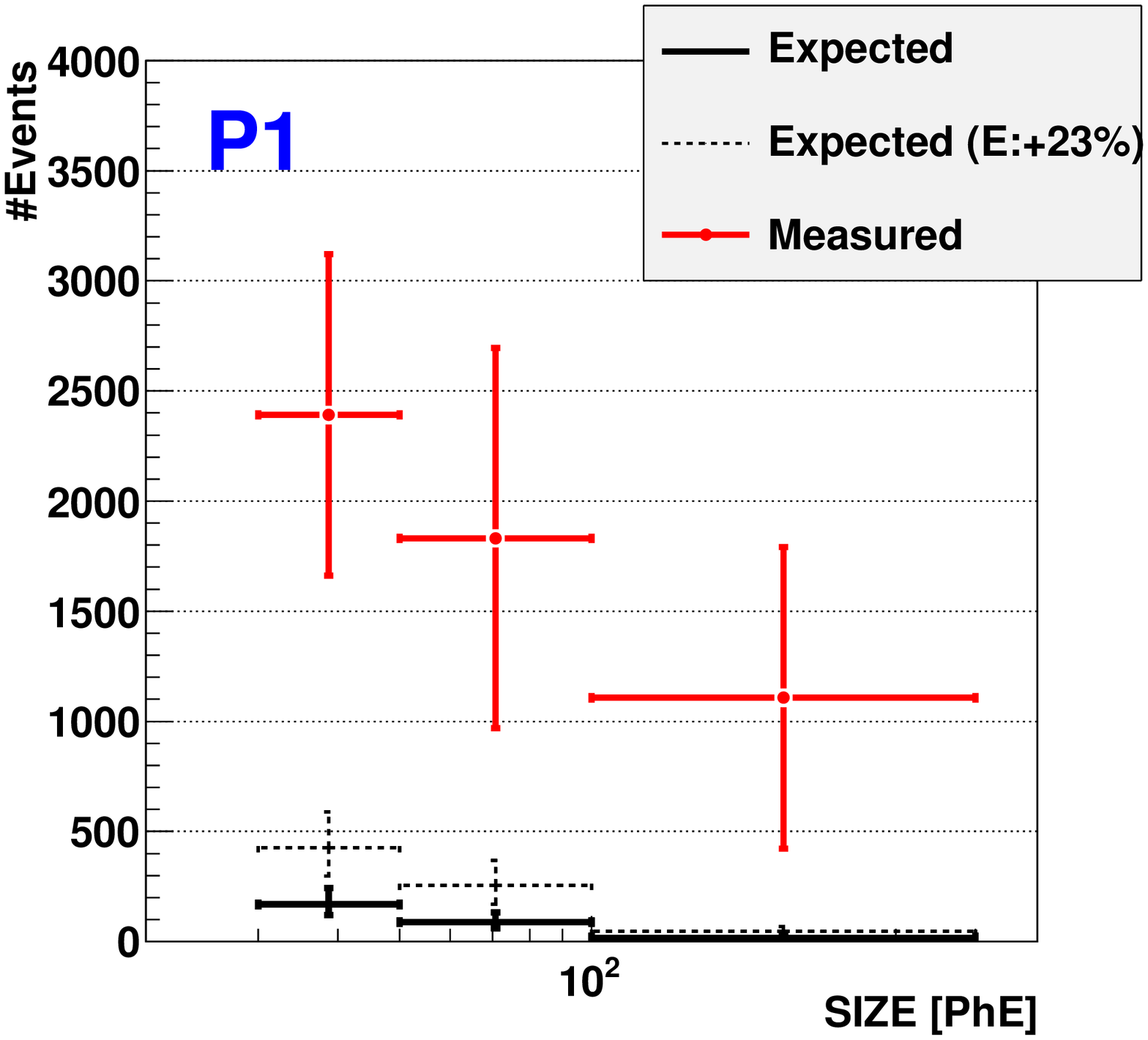}\\
\includegraphics[width=0.395\textwidth]{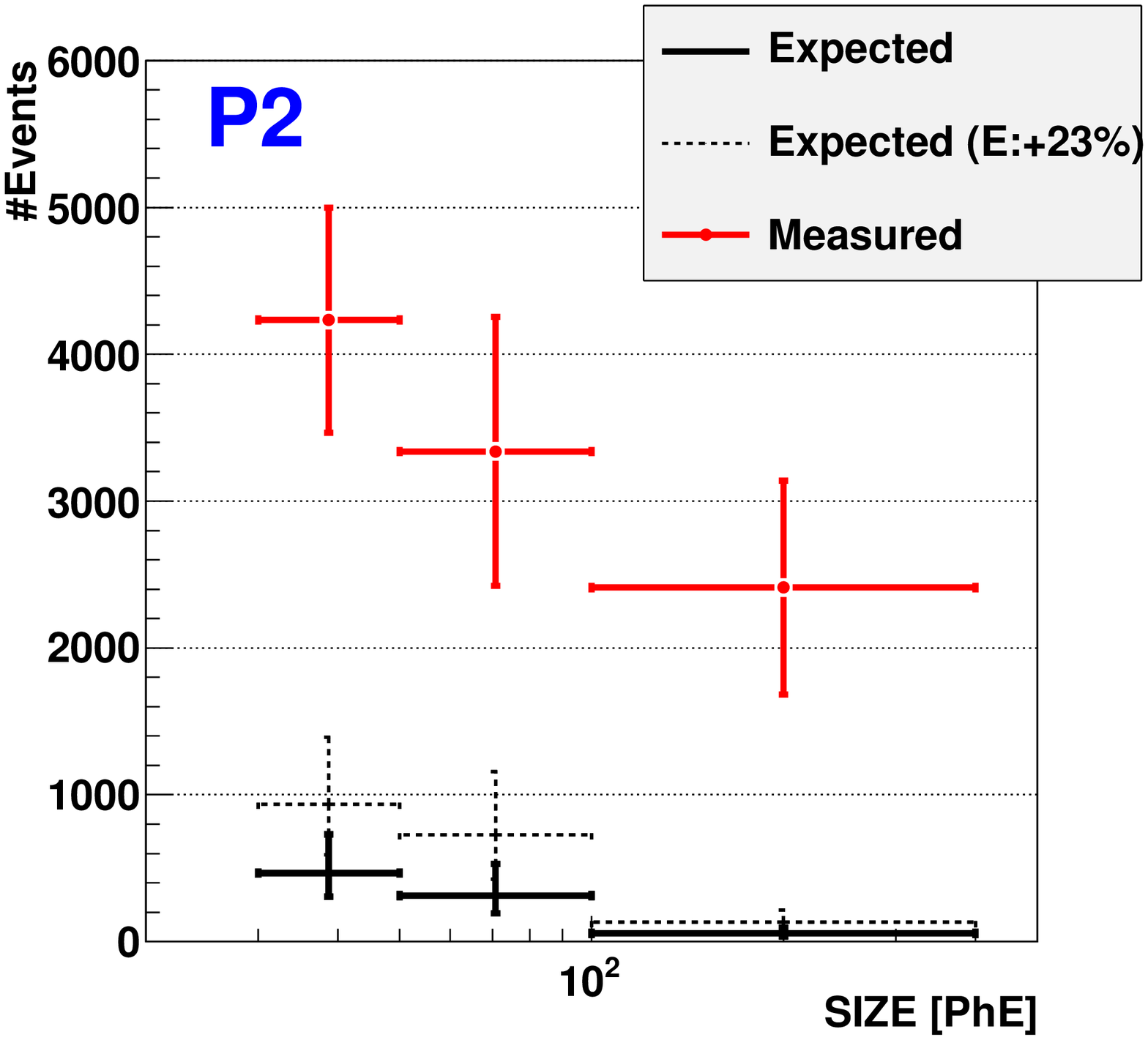}
\caption{Comparison of the $SIZE$ distribution between the expectations 
and the measurements. From the top, P1~+~P2, P1 and P2
are shown. Red lines show the measured distributions
while the solid black lines indicate the 
expected distributions computed with MC simulations 
assuming the exponential cutoff spectra determined by {\it Fermi}-LAT (see
Table \ref{TabSpec} and Figure \ref{FigSpectrumPhase}).
Dotted lines are the expectation in the case where the energy scale of MAGIC is 23\% higher than
that of {\it Fermi}-LAT due to systematic uncertainties of both instruments.
}
\label{FigIncons}
\end{figure}

\subsection{Statistical Evaluation of the Difference between the Extrapolated {\it Fermi}-LAT Spectrum and the MAGIC Data}
\label{SectIncons}
For MAGIC, the energy resolution below 50\,GeV is 
about 40\%.
In addition, near the trigger threshold, the energy is overestimated 
because only events which deposit more Cherenkov photons onto the MAGIC mirrors are
selectively triggered. These effects can be corrected with MC simulations, but
the correction introduces additional systematic uncertainties.
In order to minimize these uncertainties, we adopt the following method.
Assuming that the exponential cutoff spectrum determined by the {\it Fermi}-LAT below a few tens of GeV
is valid up to 2\,TeV, we calculate the expected $SIZE$ distribution 
in the MAGIC data. 
Taking into account the trigger threshold of 27 PhE,
the events with $SIZE$ below 30 PhE are not used to avoid a possible mismatch between
MC and real data near the threshold.
Then, we compute the $\chi^2$ value between the expected and measured distributions.
Statistical errors of $E_c$ measured by {\it Fermi}-LAT are 
taken into account as an error of the expected distribution.

The results are shown in Figure \ref{FigIncons}.
The $\chi^2$ values are 54.2 , 15.8 and  42.3
for P1~+~P2, P1 and P2, respectively.
The number of degrees of freedom is 3.
The exclusion probabilities correspond to 6.7$\sigma$,
3.0$\sigma$ and 5.8$\sigma$. It should be noted that the possible energy scale shift
between the two instruments is not taken into account here.

The systematic uncertainty in the energy scale of the {\it Fermi}-LAT 
is estimated to be less than 7\% above 1\,GeV \citep{FermiDiffuse}
while that of MAGIC is estimated to be 16\% \citep{MAGICCrab}. 
We performed 
the same statistical test with an increased {\it Fermi}-LAT energy scale of 23\%,
corresponding to the linear sum of the systematic errors in both instruments.
The results are shown as dotted lines in Figure \ref{FigIncons}.
Though the discrepancies become smaller, 
the $\chi^2$ values are 42.3, 12.6 and 30.0
with the number of degree of freedom 3 for P1~+~P2, P1 and P2, respectively. 
The exclusion probabilities correspond to 5.8$\sigma$, 2.5$\sigma$ and 4.7$\sigma$. 
Even with the systematic uncertainties taken into account, 
the inconsistency between the extrapolated {\it Fermi}-LAT spectrum and the observations
by MAGIC is significant.

\section{Energy spectra}

In the previous section, it was shown that 
the extrapolation of the {\it Fermi}-LAT measured spectra under the exponential cutoff assumption 
results in significant differences with the MAGIC data above 25\,GeV.
Here we present the energy spectrum between 25\,GeV and 100\,GeV
based on the MAGIC measurements, which are the first flux measurements
in this energy region and complement the {\it Fermi}-LAT and VERITAS measurements \citep{VeritasArxiv}.
\begin{figure}[h]
\centering
\includegraphics[width=0.45\textwidth]{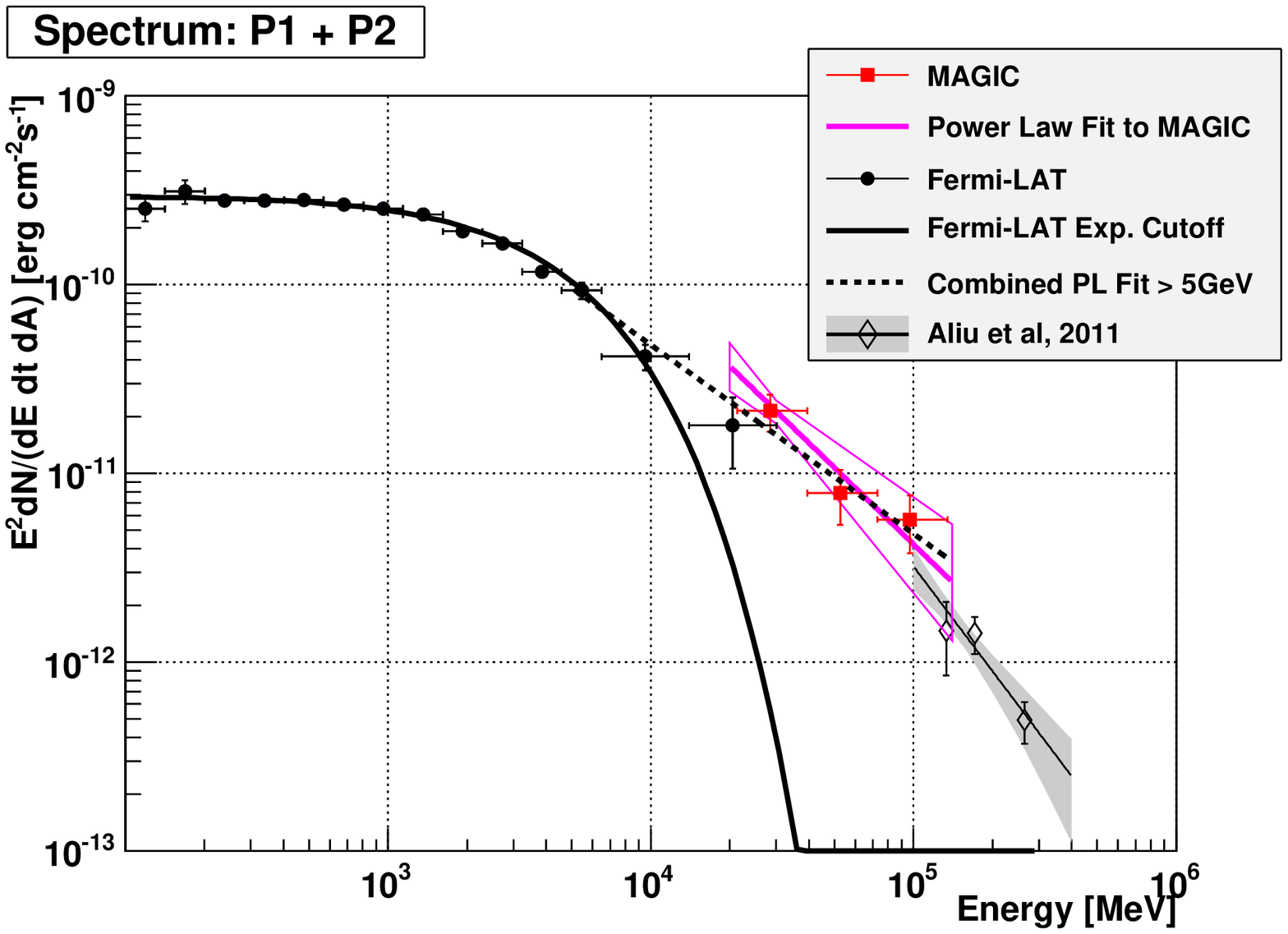}\\
\includegraphics[width=0.45\textwidth]{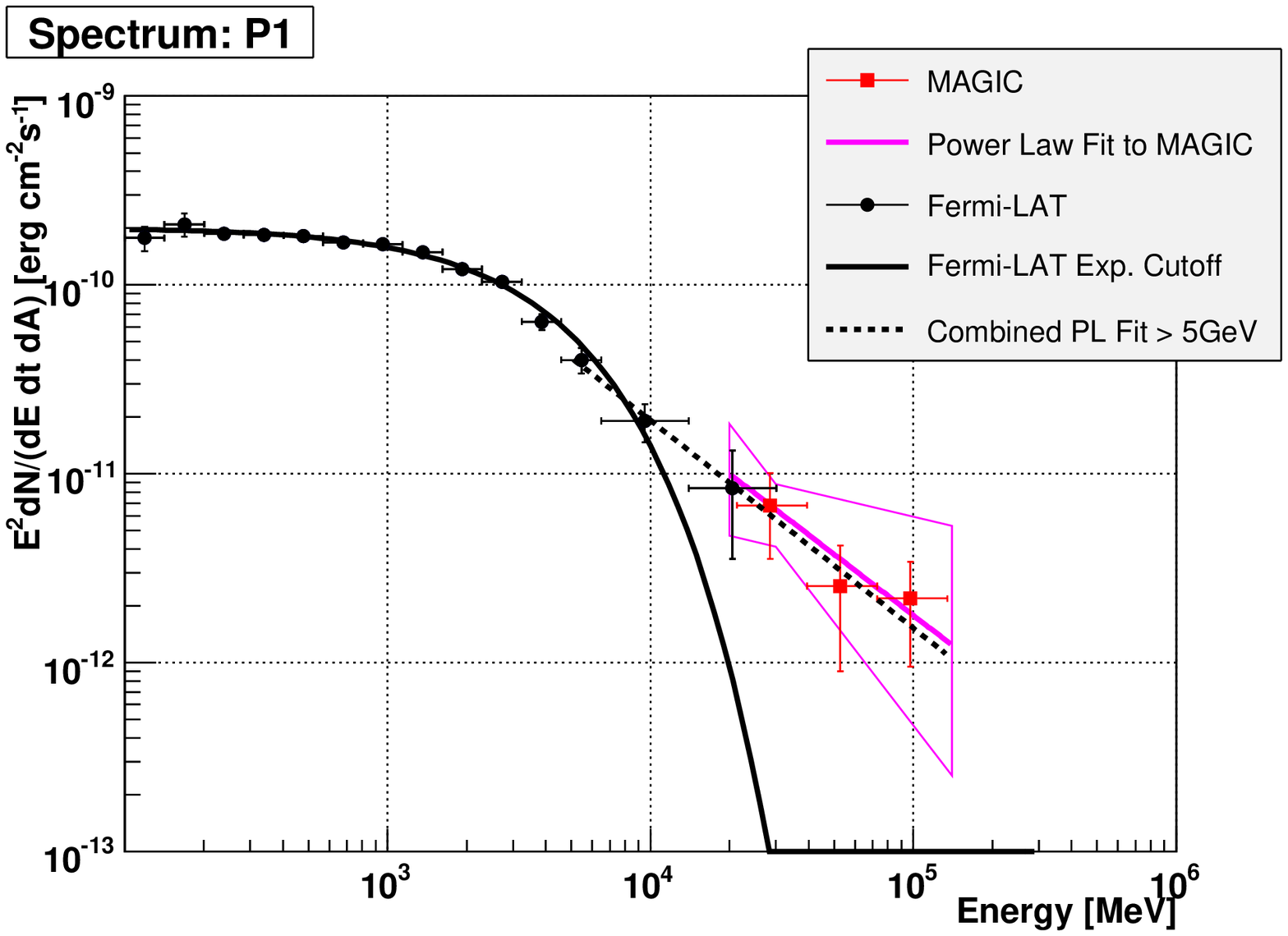}\\
\includegraphics[width=0.45\textwidth]{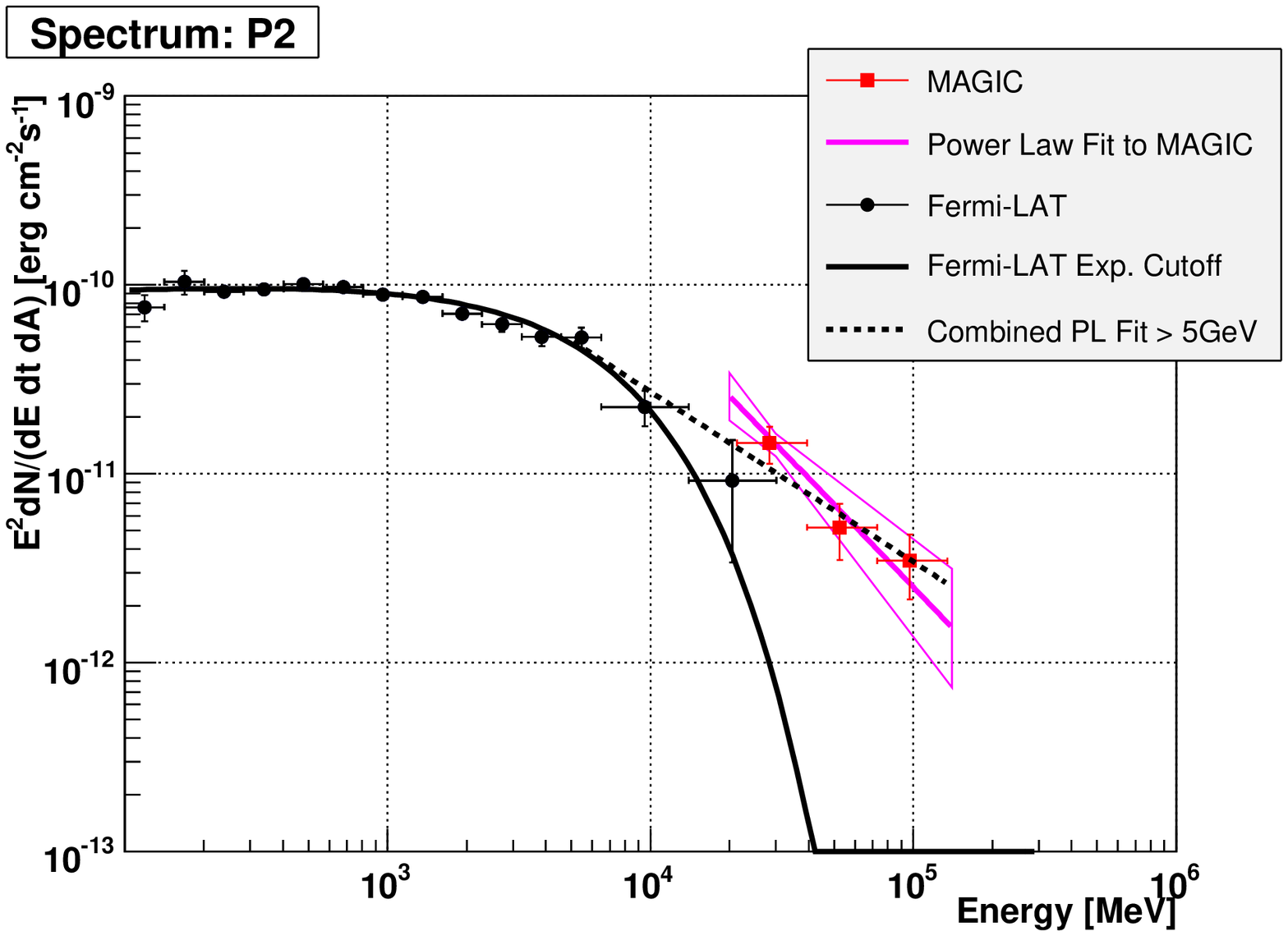}
\caption{Energy spectra of the Crab pulsar
for P1~+~P2, P1, and P2 from the top. 
The black solid lines and dots are obtained from the public {\it Fermi}-LAT data,
while red points denote the MAGIC measurements. The pink line and a butterfly shape
indicate the power-law fit to the MAGIC data and its statistical uncertainty.
The gray shade with a black line and open diamonds denote the VERITAS measurements \citep{VeritasArxiv}.
The dotted lines show the results of the combined fit above 5\,GeV (see Section \ref{SectCombFit}).
}
\label{FigSpectrumPhase}
\end{figure}

\begin{table*}
\scriptsize
\begin{center}
\caption{The best fit parameters of the spectra for different phase intervals.\label{TabSpec}
}
\begin{tabular}{|r|r|r|r|r|r|}
\tableline
 & \multicolumn{3}{|c|}{{\it Fermi}-LAT\tablenotemark{a}} & \multicolumn{2}{|c|}{MAGIC\tablenotemark{b}}  \\
\tableline
Phase & $F_1$  [10$^{-10}$cm$^{-2}$s$^{-1}$MeV$^{-1}]$ & $\Gamma_1$ & $E_c$ & $F_{30}$ [10$^{-9}$cm$^{-2}$s$^{-1}$TeV$^{-1}$]  & $\Gamma_2$ \\
\tableline
\tableline
Total & $2.32 \pm 0.05_{stat} $ & $1.99 \pm 0.02_{stat}$ & $6.1\pm 0.5_{stat}$ & & \\
\tableline
P1~+~P2 & $1.94 \pm 0.05_{stat}$ & $1.98 \pm 0.02_{stat}$ & $4.5 \pm 0.3_{stat}$ & $14.9 \pm 2.9_{stat} \pm 9.6_{syst}$ & $3.4 \pm 0.5_{stat} \pm 0.3_{syst} $\\
\tableline	
P1      & $1.29 \pm 0.04_{stat}$ & $1.99 \pm 0.02_{stat}$ & $3.7 \pm 0.3_{stat}$ &  $4.5 \pm 2.3_{stat} \pm 2.6_{syst}$& $3.1 \pm 1.0_{stat} \pm 0.3_{syst}$\\
 \tableline
P2      & $0.67 \pm 0.02_{stat}$ & $1.95 \pm 0.03_{stat}$ & $5.9 \pm 0.7_{stat}$ & $ 10.0 \pm 1.9_{stat} \pm 6.7_{syst}$ & $3.4 \pm 0.5_{stat} \pm 0.3_{syst}$\\\tableline
\end{tabular}
\end{center}
\tablenotetext{a}{Obtained by fitting Equation (\ref{EqExpCut}) to {\it Fermi}-LAT data.}
\tablenotetext{b}{Obtained by fitting Equation (\ref{EqPLMAGIC}) to MAGIC data.}
\end{table*}

\subsection{Spectra of P1, P2 and P1~+~P2}
The energy spectra of the Crab pulsar were computed based on the detected excess events
found in P1 and P2, using the standard MAGIC software.
The energy resolution and the trigger bias effect were corrected by an unfolding procedure
which includes the Tikhonov regularization method \citep{Tikhonov}.
The results are shown in Figure \ref{FigSpectrumPhase}.
The combined spectrum of P1~+~P2 is consistent with a power law, which can be described using the
following formula:
\begin{eqnarray}
\frac{{\rm d}^3 F(E)}{{\rm d}E {\rm d}A{\rm d}t} = F_{30} (E/30{\rm ~GeV})^{-\Gamma_2} 
\label{EqPLMAGIC}
\end{eqnarray}
 $F_{30} = (14.9 \pm 2.9_{stat} \pm 9.6_{syst}) \times 10^{-9}~{\rm cm^{-2}s^{-1} TeV^{-1}}$
and $\Gamma_2~=~{3.4~\pm~0.5_{stat} \pm 0.3_{syst}}$ were obtained as best-fit parameters. Also, the spectrum seems to 
connect smoothly to the VERITAS measurements above 100\,GeV \citep{VeritasArxiv}.

The individual spectra of P1 and P2 were also calculated
using the same data set. 
Results are shown as well in Figure \ref{FigSpectrumPhase}.
The best-fit parameters are summarized in Table \ref{TabSpec}.

Figure \ref{FigSpectrumPhase} clearly shows the deviation of the MAGIC
spectra with respect to the extrapolation
of the exponential cutoff spectra determined by {\it Fermi}-LAT, which
is consistent with our statistical analysis in the previous section.

\subsection{Combined Fit above 5\,GeV}
\label{SectCombFit}
To get a better estimate of the power-law index for the higher energies, {\it Fermi}-LAT data
points above 5\,GeV and MAGIC data points are combined and fitted by a power law.

\begin{eqnarray} 
\frac{{\rm d}^3F(E)}{{\rm d}E{\rm d}A {\rm d}t} &=& F_{10} (E/10{\rm GeV})^{-\Gamma} \nonumber \\
\label{EqScaling}
\end{eqnarray} 

It should be mentioned that {\it Fermi}-LAT points are obtained using
the likelihood analysis for each energy interval assuming a power law, while
for the fit each point was assumed to follow 
Gaussian statistics with the standard deviation being the error obtained by the likelihood results.
Though this does not statistically correspond to the exact likelihood function of the problem, it is a good and appropriate approximation.
The results are shown in Table \ref{TabCombFitPL}.
　

~\\
~\\
\section{A closer look at the Pulse Profiles}
\label{SectProfile} 
\subsection{Peak Phase and Pulse Width}
We examine the peak phase and the pulse width in the MAGIC energy range
assuming a pulse shape a priori and fitting it to the measured profile.
The used functions are a Gaussian
\begin{eqnarray}
F_G(x) = F_0 \exp(- \frac{(x-\mu)^2}{2\sigma^2})
\label{EqGaus}
\end{eqnarray}
and a Lorentzian
\begin{eqnarray}
F_L(x) = F_0 \left(1+ \frac{(x-\mu)^2}{\sigma^2} \right)^{-1},
\label{EqLorentz}
\end{eqnarray}
where
$\mu$ corresponds to the peak phase, while $\sigma$ can be translated into a pulse width.
The full width at half-maximum (FWHM) is equal to $2.35 \times \sigma$ and
$2 \times \sigma$ for the Gaussian and the Lorentzian, respectively.
In order to study the asymmetry of the pulses, 
asymmetric Gaussian and Lorentzian functions are also assumed with $\sigma$ 
being different below and above $x=\mu$.

\begin{figure*}[t]
\centering
\includegraphics[width=0.38\textwidth]{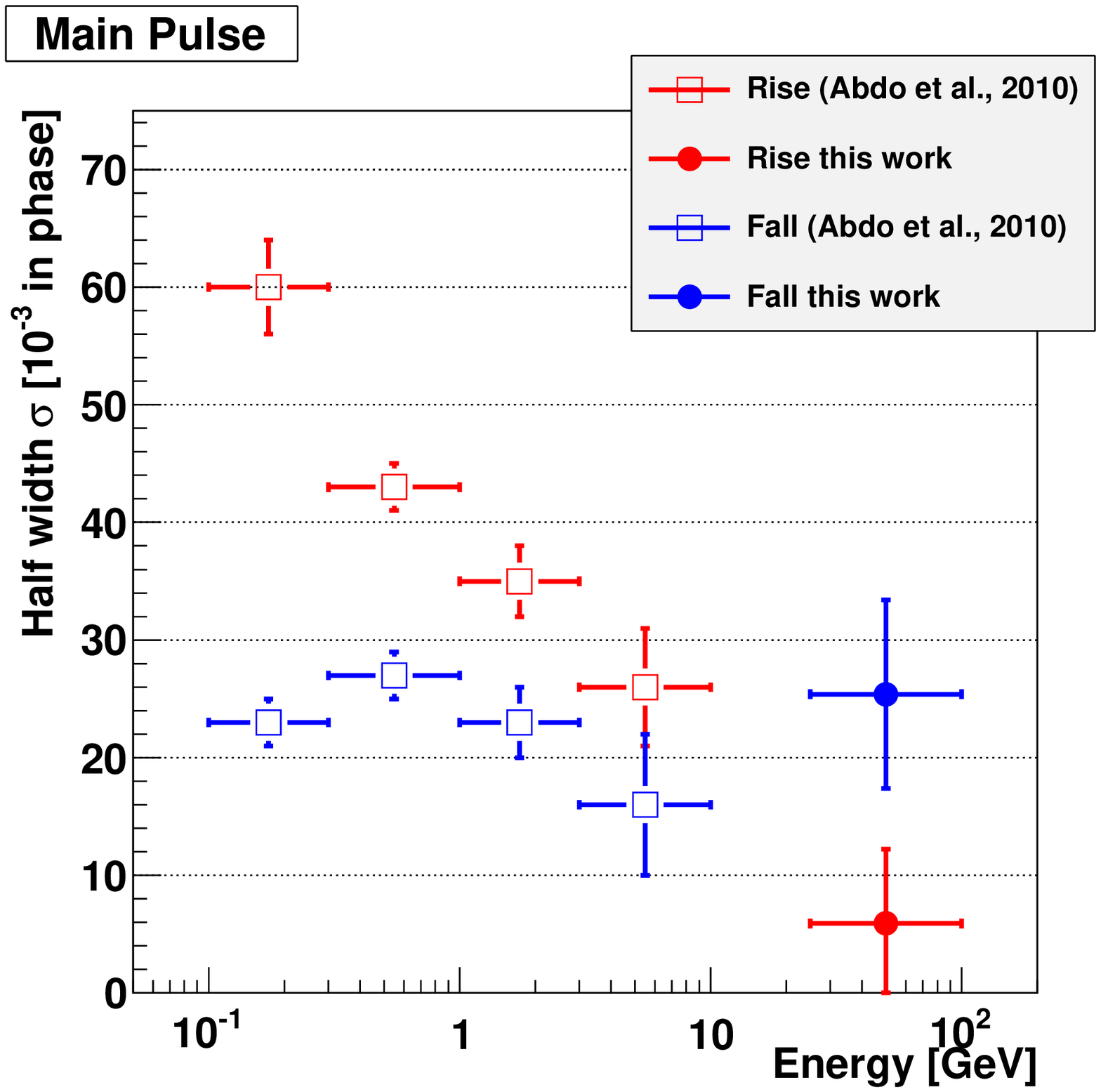}
\includegraphics[width=0.38\textwidth]{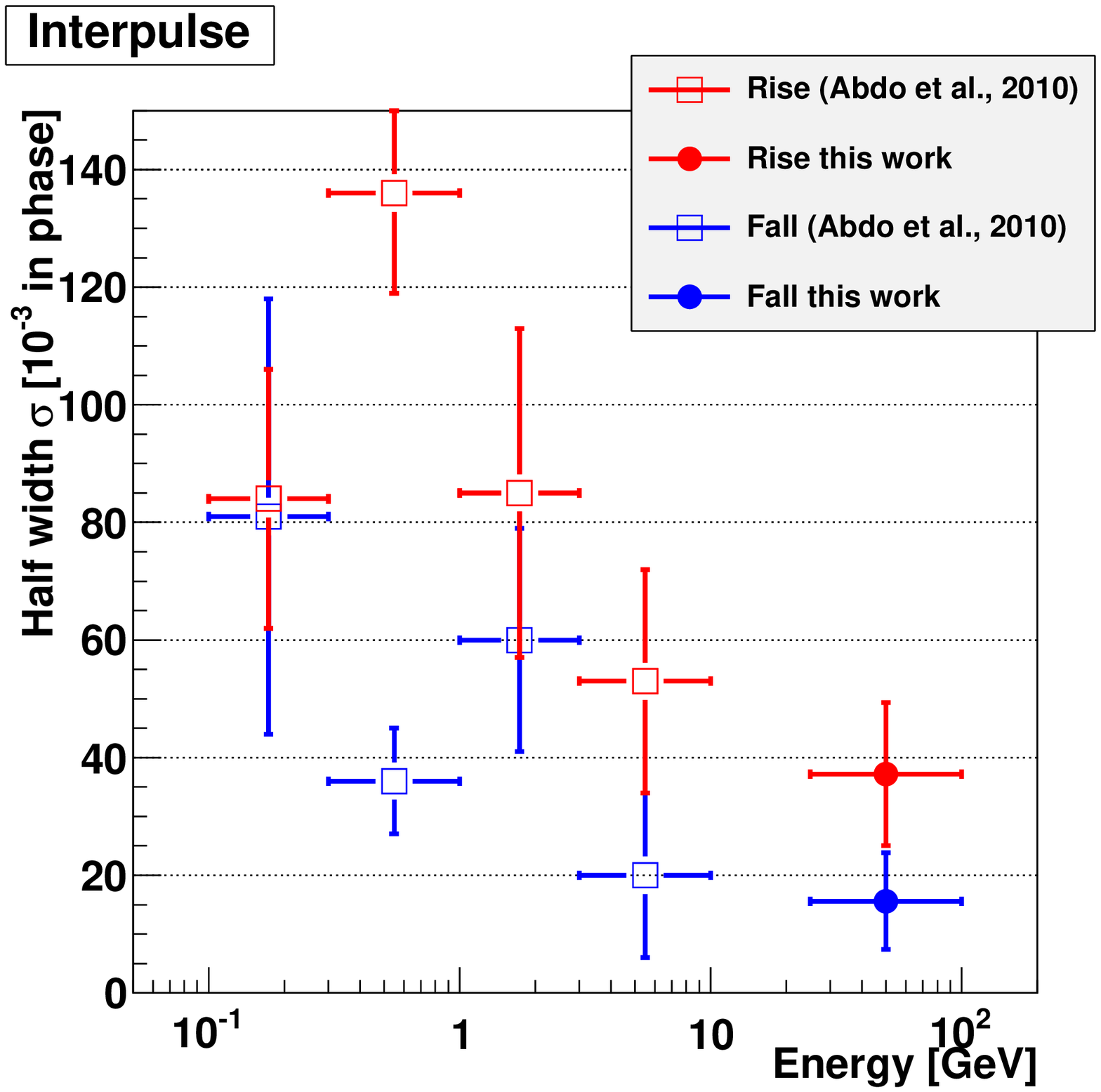}
\caption{Energy dependence of the rising and falling half-width
for the main pulse (left) and the interpulse (right)
assuming an asymmetric Lorentzian as the pulse shape. 
Points below 10\,GeV are reproduced from \citet{FermiCrab}.
}
\label{FigWidth}
\end{figure*}

The results are shown in Table \ref{TabShapes}.
The peak phase of the main pulse is compatible with 0.0 (defined by the radio peak) 
for all fitted parameterizations.
The {\it Fermi}-LAT Collaboration reported that the pulse shape is well modeled by an asymmetric Lorentzian 
and the peak phase above 100\,MeV is $-0.0085 \pm 0.0005$ \citep{FermiCrab}.
The MAGIC result under the asymmetric Lorentzian assumption is consistent with it as well.
The peak phase of the interpulse depends on the assumption of the shape, while 
it is approximately 0.39 $\pm$ 0.01, which is also consistent with the value
 above 100\,MeV ($0.398 \pm 0.003$, \citet{FermiCrab}).
The FWHM of the main pulse is approximately $0.03 \pm 0.01$ independently of 
the assumed shape, while that of the interpulse
is $0.07 \pm 0.01$ and $0.05 \pm 0.01$ for the assumption of the Gaussian 
shaped and Lorentzian shaped pulse, respectively.
The main peak is narrower than the interpulse.
The asymmetric assumptions imply that, for the main pulse, the rising edge
is steeper than the falling edge, while the opposite is true for the interpulse. 
In Figure \ref{FigWidth}, the half-widths $\sigma$ for the rising and falling edges 
of both the main pulse and the interpulse are compared with the values reported in \citet{FermiCrab}. 
The rising half of the main pulse become narrower as the energy increases.
Though the uncertainty is larger, a similar tendency is also visible in the rising and falling halves of the interpulse, while no such energy dependence is visible in the falling half
of the main pulse.

\begin{table}[t]
\centering
 \caption{The power law fit (Equation (\ref{EqScaling})) combining the {\it Fermi}-LAT data above 5\,GeV and the MAGIC data.
\label{TabCombFitPL}}
 \begin{tabular}{|c|c|c|c|}
\tableline
Phase & $F_{10}$ [10$^{-7}$ cm$^{-2}$s$^{-1}$TeV$^{-1}$] & $\Gamma$ & $\chi^2/n.d.f$\\
\tableline
P1~+~P2 & 3.0 $\pm$ 0.2 & 3.0 $\pm$ 0.1 & 8.1/4\\
P1 & 1.2 $\pm$ 0.2 & 3.1 $\pm$ 0.2 & 1.6/4\\
P2 & 1.7 $\pm$ 0.2 & 2.9 $\pm$ 0.1 & 8.3/4\\
\tableline
\end{tabular}
\end{table}

\begin{figure*}[t]
\centering
\includegraphics[width=0.7\textwidth]{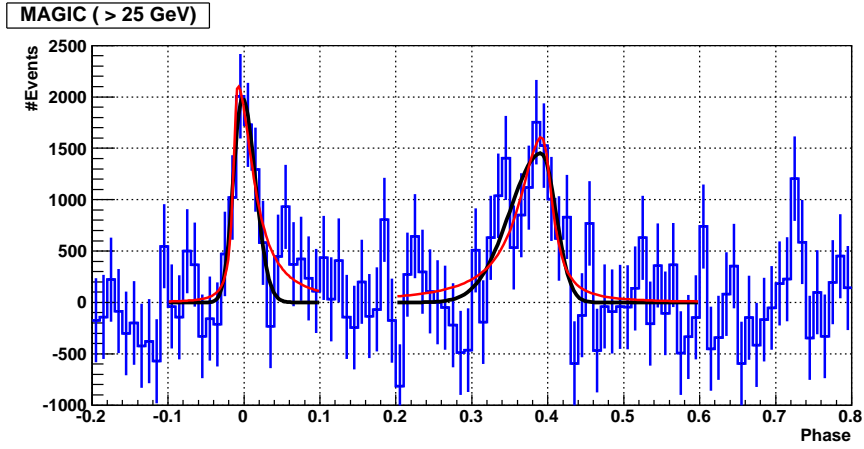}
\caption{Pulse profile of the MAGIC data fitted by asymmetric Lorentzians (red lines)
and asymmetric Gaussians (black lines).
}
\label{FigCloserLook}
\end{figure*}

\begin{table*}[t]
\centering
\caption{Fit results using different assumptions on the shape of the main pulse and the interpulse.\label{TabShapes}
}
{\renewcommand \arraystretch{1.0}
 \begin{tabular}{|c|c|c|c|c|c|}
\tableline
\tableline
 & \multicolumn{5}{|c|}{Main pulse} \\ 
\tableline
Assumed Shape & $\mu$ & $\sigma_1$ & $\sigma_2$ & FWHM & $\chi^2$/n.d.f\\
\tableline
Gaussian & $ 0.7 \pm 2.6 $ & $ 14.2 \pm 2.4 $ & & $33.3 \pm 5.7 $ & 13.6/17\\
 \tableline
Asym. Gauss.& $ -2.2 \pm 6.6 $ & $ 11.8 \pm 5.4 $ & $ 16.9 \pm 7.0 $ & $33.7 \pm 10 $ & 13.3/16\\
 \tableline
Lorentzian & $ 0.4 \pm 2.7 $ & $ 14.4 \pm 3.8 $ & & $28.8 \pm 7.6 $ & 13.4/17\\
 \tableline
Asym. Lorentz.& $ -8.9 \pm 6.5 $ & $ 5.9 \pm 6.3 $ & $ 25.4 \pm 8.0 $ & $31.3 \pm 10 $ & 11.2/16\\
 \tableline
\tableline
& \multicolumn{5}{|c|}{Interpulse} \\
\tableline
Assumed Shape & $\mu$ & $\sigma_1$ & $\sigma_2$ & FWHM  & $\chi^2$/n.d.f\\
\tableline
Gaussian & $ 377.2 \pm 5.7 $ & $ 32.4 \pm 4.8 $ & & $76 \pm 11 $ & 38.5/37\\
 \tableline
Asym. Gauss.& $ 391.8 \pm 10 $ & $ 42.1 \pm 9.1 $ & $ 18.8 \pm 8.2 $ & $72 \pm 14 $ & 36.6/36\\
 \tableline
Lorentzian & $ 384.1 \pm 5.2 $ & $ 26.2 \pm 7.2 $ & & $52 \pm 14$ & 41.6/37\\
 \tableline
Asym. Lorentz.& $ 392.9 \pm 8.7 $ & $ 37.2 \pm 12.2 $ & $ 15.6 \pm 8.2 $ & $53 \pm 15 $ & 39.5/36\\
 \tableline
\tableline
 \end{tabular}
}
\tablenotetext{*}{Units of all the parameters are $10^{-3}$ in phase}
 \end{table*}

\subsection{Other Emission Components}

The AGILE Collaboration reported a possible third peak 
at phase between 0.65 and 0.8 above 100 MeV with the significance of 3.7$\sigma$
\citep{Pellizzoni2009}.
A hint of a third peak is also seen in the {\it Fermi}-LAT data at phase $\sim 0.74$ only above 10\,GeV with the significance of 2.3$\sigma$
\citep{FermiCrab}.
They coincide with the radio peak observed between 4.7 and 8.4\,GHz. 
In the MAGIC data above 25\,GeV, a similar peak is seen at phase $\sim 0.73$.
Defining the signal phases as $0.72 - 0.75$ 
and the background control phases as the OP phases ($0.52 - 0.88$) excluding the signal phases,  
$1600 \pm 700$ excess events were found, corresponding to 2.2$\sigma$.
This pre-trial significance is too low to claim a detection
and it is within the range of expected fluctuation of the background.

An emission between the main pulse and the interpulse, i.e., a so-called bridge emission 
is seen in some energy bands. With the MAGIC data above 25\,GeV, it is not visible though 
the statistical uncertainty is large. 
Defining the signal phases as $0.04 - 0.32$, i.e., from the end of P1 to the beginning of P2,
and using the OP phases as the background estimate, $3200 \pm 2800$ excess events were found,
corresponding to 1.1$\sigma$. The upper limit on the number of excess events with the 95\% confidence level
is 8800, which corresponds to half of the flux of P1~+~P2.

\clearpage
\section{Discussion and Conclusions}

\subsection{Summary of Findings}

The findings of this study can be summarized as follows.

\begin{enumerate}

\item 59 hr of MAGIC observations of the Crab pulsar during the winters 2007/2008
and 2008/2009 resulted in the detection of $6200 \pm 1400$ and $11300 \pm 1500$ excess events from
P1 and P2, respectively. The flux of P2 is a factor $\sim 2$ larger than that of P1 above 25\,GeV.
\item No yearly variability in the pulse profile or in the flux was found.
\item The flux measured with MAGIC is significantly higher than the extrapolation of the
 exponential cutoff spectrum determined by {\it Fermi}-LAT.
\item The energy spectra extend up to at least 100\,GeV
and can be described by a power law between 25\,GeV and 100\,GeV.
The power-law indices of P1, P2 and P1~+~P2 are
$-3.4 \pm 0.5_{stat} \pm 0.3_{syst}$,
$-3.1 \pm 1.0_{stat} \pm 0.3_{syst}$ and
$-3.4 \pm 0.5_{stat} \pm 0.3_{syst}$, respectively.
The sensitivity of MAGIC I above 100\,GeV is not sufficient to clarify
if the spectrum continues with a power law or drops more rapidly.
However, our spectrum of P1~+~P2 and the VERITAS measured spectrum above 100\,GeV
seem to be a good extrapolation of each other.
\item Assuming an asymmetric Lorentzian for the pulse shape, 
the peak positions of the main pulse and the interpulse are estimated to be
$-0.0089 \pm 0.0065$ and  $0.3929 \pm 0.0087$ in phase, while the FWHMs
are $0.031 \pm 0.010$ and  $0.053 \pm 0.015$.
Compared with the {\it Fermi}-LAT measurements, the pulse widths are narrower in the MAGIC energy regime.
\item  The bridge emission between P1 and P2 is weak. With the current sensitivity
no signal was found. A potential third peak with a pre-trial significance of 2.2$\sigma$
is seen at a similar position as in the AGILE data above 100 MeV and
the {\it Fermi}-LAT data above 10\,GeV, 
but it is consistent with the background fluctuation.
\end{enumerate}

The spectrum of the Crab pulsar does not follow an exponential
cutoff but, after the break, it continues as a power law. 
This is inconsistent with the OG
and SG models in their simplest version, where it is assumed
that the emission above $\sim 1$\,GeV comes only from curvature radiation,
 leading to an exponential cutoff in the spectrum. \\
A theoretical interpretation of this deviation from the exponential cutoff
is discussed in the next section.

\subsection{Theoretical Interpretation of the Spectrum}

In a pulsar magnetosphere, 
high-energy photons are emitted by the electrons and positrons 
that are accelerated by the magnetic-field-aligned electric field,
$\mbox{\boldmath $E_\parallel$}$.
To derive $\mbox{\boldmath $E_\parallel$}$, 
we must solve the inhomogeneous part of the Maxwell equations
\citep{Fawley1977, Scharlemann1978, Arons1979}
\begin{equation}
  \nabla\cdot\mbox{\boldmath $E_\parallel$}
  = 4\pi (\rho-\rho_{\rm GJ})
  \label{eq:PoissonEQ}
\end{equation}
where $\rho$ denotes the real charge density, and
$\rho_{\rm GJ} \equiv
  -\mbox{\boldmath $\Omega$}\cdot\mbox{\boldmath $B$}/(2\pi c)$
the Goldreich$-$Julian charge density
\citep{GJ, Mestel1971};
the angular-velocity vector $\mbox{\boldmath $\Omega$}$ 
points in the direction of the NS spin axis with magnitude
$\vert \mbox{\boldmath $\Omega$} \vert=2\pi/P$,
$\mbox{\boldmath $B$}$ refers to the magnetic field, 
and $c$ is the speed of light. If $\rho$ coincides with $\rho_{\rm GJ}$ in the entire magnetosphere,
$\Ell$ vanishes everywhere.
However, if $\rho$ deviates from $\rho_{\rm GJ}$ in some region,
it is inevitable for a non-vanishing $\Ell$ to arise around the region
and the particle accelerator, or the so-called the ``gap'', appears.

To predict the absolute luminosity of the gap,
as well as any phase-averaged and phase-resolved emission properties,
we must constrain 
$E_\parallel \equiv \vert\mbox{\boldmath $E_\parallel$}\vert$, 
$\rho$, and the gap geometry
in the three-dimensional magnetosphere.
We solve the Poisson equation~(\ref{eq:PoissonEQ})
together with the Boltzmann equation for electrons and positrons
($e^\pm$'s) and with the radiative transfer equation 
between $0.005$~eV and $10$~TeV.
The position of the gap is solved within the free-boundary framework
so that $\Ell$ may vanish on the boundaries,
and turned out to distribute in the higher altitudes
as a quantitative extension of previous OG models,
which assume a vacuum (i.e., $\rho=0$) in the gap.
In the present paper, we propose a new, non-vacuum (i.e., $\rho\ne 0$) 
OG model, solving the distribution of $\rho$ self-consistently from the
$e^-$--$e^+$ pair creation at each point 
\citep{Beskin1992, Hirotani1998, Hirotani1999a, Hirotani1999b, Hirotani1999c, Takata2006}.

The created $e^\pm$'s are polarized and accelerated by $\Ell$ in the gap 
to attain high Lorentz factors, $\sim 10^{7.5}$.
Such ultra-relativistic $e^+$'s and $e^-$'s emit 
primary $\gamma$-rays via synchro-curvature and inverse Compton (IC)
processes.
For the IC process, the target photons are emitted from the
cooling NS surface, from the heated PC surface,
and from the magnetosphere in which pairs are created.
The primary $\gamma$-rays that are emitted by the 
(inwardly accelerated) $e^-$'s
efficiently (nearly head-on) collide with the surface X-rays
to materialize as the primary $e^\pm$'s within the gap
and as the secondary $e^\pm$'s outside the gap.
The secondary $e^\pm$'s efficiently lose their energy via the synchrotron
process in the inner magnetosphere and cascade into tertiary and
higher-generation $e^\pm$'s via two-photon and one-photon
(i.e., magnetic) pair-creation processes.
The primary $\gamma$-rays that are emitted by the 
(outwardly accelerated) $e^+$'s
via the curvature process collide with the magnetospheric X-rays
to materialize as secondary $e^\pm$'s with initial Lorentz factors
$\gamma \sim 10^{3.5}$ outside the gap, 
while those emitted via the IC process 
collide with the magnetospheric IR$-$UV photons
to materialize with $\gamma \sim 10^7$.
The secondary pairs with $\gamma \sim 10^{3.5}$
emit synchrotron emission below 10~MeV and 
synchrotron-self-Compton (SSC) emission between 10~MeV and a few GeV
(by up scattering the magnetospheric X-ray photons).
The secondary pairs with $\gamma \sim 10^7$, on the other hand,
emit synchrotron emission below 10~GeV and
SSC emission between 10~GeV and 1~TeV
(by up-scattering the magnetospheric IR$-$UV photons).
Such secondary SSC photons (between 10~GeV and 1~TeV)
are efficiently absorbed colliding with 
the magnetospheric IR$-$UV photons 
to materialize as tertiary pairs with $10^4 < \gamma < 10^6$.
In the high-energy end (i.e., $10^5 < \gamma < 10^6$),
the tertiary pairs up-scatter magnetospheric IR$-$UV photons
into the energy range between 1~GeV and 300 GeV
Therefore, if we focus the emission component above 10~GeV,
the photon are emitted by the secondary and tertiary pairs
when they up-scatter the magnetospheric IR$-$UV photons
(see also \citet{Lyutikov2011} for an analytical discussion 
 of this process using the multiplicity factor of higher-generation pairs).
Note that the primary IC photons are emitted from relatively
inner part of the outer magnetosphere and hence totally absorbed
by the magnetospheric IR-UV photon field,
of which specific intensity is self-consistently solved together
with $\Ell$ and the particle distribution functions at each point.
Note also that the surface X-ray field little 
affects the pair creation or the IC process in the outer magnetosphere
of the Crab pulsar.
For the details of this self-consistent approach,
see \citet{Hirotani2006}.

\begin{figure}
 \includegraphics{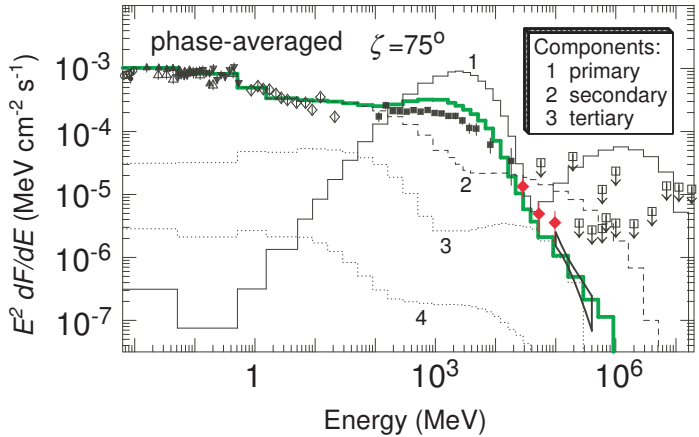}
\caption{
Phase-averaged spectrum 
of the pulsed emission from the Crab pulsar
predicted by the self-consistent outer Gap model. 
The thin solid line (labeled ``1'') 
represents the flux of the photons emitted by
the primary positrons accelerated in the gap,
while the thin dashed one (labeled ``2'') 
and the thin dotted one (labeled ``3'') 
by the secondary and tertiary pairs,
respectively, created outside the gap.
The thick green solid line includes magnetospheric absorption 
and subsequent reprocesses, and hence represents the flux
to be observed.
Interstellar absorption is not considered. 
The filled circles (LECS), 
open circles (MECS), filled triangles (PDS) denote the BeppoSAX 
observations, while the open triangles the Gamma-ray Imaging 
Spectrometer (GRIS). 
Inverse filled triangles (OSSE), open diamonds (COMPTEL) 
denote CGRO observations.
The filled squares denote the {\it Fermi}-LAT observations, while
the red filled diamonds the MAGIC observations (this work).
The butterfly shape above 100 GeV indicates the VERITAS observations \citep{VeritasArxiv}.
The ordinate is in 
$\mbox{(MeV)}^2\,\mbox{s}^{-1}\,\mbox{cm}^{-2}\,\mbox{MeV}^{-1}$ unit. 
Data points are from \citet{Kuiper2001} and \citet{FermiCrab}
\label{fig:Crab_OG}
}
\end{figure}

In Figure~\ref{fig:Crab_OG},
we present the solved spectral energy distribution 
of the total pulse component
when the magnetic axis is inclined $60^\circ$ with respect to the
rotational axis and when the observer's viewing angle is $75^\circ$.
The temperature of the cooling NS emission 
is assumed to be 70~eV.
The distance to the pulsar is assumed to be 2~kpc.
The thick, green solid line represents the spectrum to be observed
(i.e., with magnetospheric absorption),
while the thin black solid line
indicates the photons emitted by the primary positrons
(with Lorentz factor $\sim 10^{7.5}$ inside the gap),
and the thin dashed line does those emitted by the secondary pairs
(with Lorentz factor $\sim 10^{3.5}$ or $10^7$ 
 outside the gap). 
Interstellar absorption is not taken into account.
The primary, un-absorbed IC component becomes
prominent above 40~GeV; however, most of such photons are
absorbed by two-photon pair production and reprocessed in lower energies 
as the secondary SSC component.
In the secondary emission component, which is depicted by the dashed curve,
there is a transition of the dominant component: the synchrotron component
dominates the IC component (due to the SSC process) below a few GeV,
whereas the latter dominates the former above this energy. 

It should be noted that the pulsed emission between 25~GeV and 180~GeV
is dominated by the SSC component emitted by the secondary and
tertiary $e^\pm$'s.
Since the higher-generation components,
which are denoted by the dotted curves,
are emitted from the higher altitudes
(near the light cylinder),
they are less efficiently absorbed,
thereby appearing as pulsed flux above 20~GeV.
The resultant spectrum (green, thick solid line)
exhibits a power-law-like shape above 20~GeV,
rather than an exponential cutoff.
We, therefore, interpret that the detected $\gamma$-rays above 25~GeV 
are mainly emitted via the SSC process
when the secondary and tertiary pairs 
up scatter the magnetospheric synchrotron IR$-$UV photons.

Although the present theoretical result is obtained by 
simultaneously solving the set of Maxwell and Boltzmann equations
under appropriate boundary conditions,
it does not rule out other possibilities 
such as the synchrotron emission (e.g., \citet{Chkheidze2011})
or the IC emission (e.g., \citet{Bogovalov2000}) 
from the wind zone by ultra-relativistic particles,
which may be accelerated by MHD interactions or by magnetic reconnection, 
for instance.
However, the fuller study of other theoretical models lies
outside the scope of the present paper.

\subsection{Outlook}

 For further studies of the pulsar emission mechanisms, observations
with a higher sensitivity are essential. 
The stereoscopic system comprising the two MAGIC telescopes has 
a 50 GeV threshold and its sensitivity above 100 GeV is nearly three times
higher than that of MAGIC I \citep{StereoPerformance}.
The results of stereoscopic observations are given elsewhere 
\citep{CrabStereo}.

\acknowledgments

We thank the Instituto de Astrof\'{\i}sica de
Canarias for the excellent working conditions at the
Observatorio del Roque de los Muchachos in La Palma.
The support of the German BMBF and MPG, the Italian INFN, 
the Swiss National Fund SNF, and the Spanish MICINN is 
gratefully acknowledged. This work was also supported by 
the Marie Curie program, by the CPAN CSD2007-00042 and MultiDark
CSD2009-00064 projects of the Spanish Consolider-Ingenio 2010
programme, by grant DO02-353 of the Bulgarian NSF, by grant 127740 of 
the Academy of Finland, by the YIP of the Helmholtz Gemeinschaft, 
by the DFG Cluster of Excellence ``Origin and Structure of the 
Universe'', by the DFG Collaborative Research Centers SFB823/C4 and SFB876/C3,
and by the Polish MNiSzW grant 745/N-HESS-MAGIC/2010/0.
This work is also supported by 
the Formosa Program between National Science Council  
in Taiwan and Consejo Superior de Investigaciones Cientificas
in Spain administered through grant number 
NSC100-2923-M-007-001-MY3.
We acknowledge the use of public tools and data provided by the Fermi Science Support Center.
\\
This experiment would not have been carried out without the initial
proposal and engagement of the late Okkie de Jager.
This publication pays tribute to his memory.



{\it Facilities:} \facility{MAGIC}, \facility{Fermi}

\bibliography{ms}
\bibliographystyle{apj}

\end{document}